\documentclass[prb,twocolumn,longbibliography]{revtex4-2}
\usepackage{amsmath,bm,amssymb,graphicx}
\usepackage{color}

\begin{document}

\title{Bosonic realization of SU(3) chiral Haldane phases}
\author{Linpu Zhang}
\author{Junjun Xu}
\email{jxu@ustb.edu.cn}
\affiliation{Beijing Weak Magnetic Detection and Applied Engineering Technology Research Center, School of Mathematics and Physics, and Institute of Theoretical Physics, University of Science and Technology Beijing, Beijing 100083, China}

\date{\today}

\begin{abstract}
We give a bosonic realization of the SU(3) antiferromagnetic Heisenberg (AFH) chain in the alternating conjugate representation, and study its phase diagram as a function of staggered interactions and anisotropy along the $T^3$ and $T^8$ directions. Unlike the SU(2) case, we observe a chiral-reversed quantum phase transition, where each competing phase is adiabatically connected to one of the chiral Haldane phases predicted in the SU(3) AFH chain with local adjoint representation. In the vicinity of the Heisenberg point, we identify a symmetry-protected topological state that appears at the first excited energy level. We also study the spontaneous $\mathbb{Z}_3$ symmetry breaking of the system, and provide a variational wavefunction that captures the transition from the topological phase to the trivial phase. Finally, we propose an experimental realization of our bosonic model by two spin-1/2 bosons in an optical lattice.
\end{abstract}

\maketitle

\section{Introduction}
Haldane predicted that one-dimensional antiferromagnetic Heisenberg (AFH) spin chains with integer spin possess a gapped ground state, and exhibit exponentially decaying correlation functions~\cite{haldane1983continuum, haldane1983nonlinear}. The model proposed by Affleck, Kennedy, Lieb, and Tasaki (AKLT) captures the essential features of the Haldane phase~\cite{affleck1987rigorous, affleck1988valence}, that each physical spin-1 particle is constructed by projecting two virtual spin-1/2 ones onto the triplet state, and two virtual spin-1/2 particles at nearest neighbor sites form a spin singlet. This spin-singlet valence bond provides a short-range entanglement, and is precisely what gives rise to its nontrivial topological characters. For example, for an open chain, the unpaired virtual spins at the boundaries give rise to effective fractional edge modes, resulting in a four-fold degeneracy of the ground state at the thermodynamic limit.

As a pioneering work of symmetry-protected topological (SPT) phases~\cite{gu2009tensor, chen2011classification, pollmann2012symmetry, pollmann2010entanglement}, the Haldane phase has recently been generalized to AFH models with higher SU($N$) symmetries~\cite{affleck1986proof, greiter2007exact, greiter2007valence, lajko2017generalization, wamer2020generalization}. Preliminary numerical calculations by Greiter \emph{et al.} suggest a correspondence between the gapped phases and their local representations ~\cite{rachel2009spinon}, that when the total number of boxes $\lambda$ in the Young tableau of the local representation is coprime with $N$, the system remains gapless and critical, while when $\lambda$ is divisible by $N$, a Haldane gap emerges, and in the intermediate case where $\lambda$ and $N$ share a nontrivial greatest common divisor $q$ with $1 < q < N$, a finite gap arises only if interactions extend to at least the $N/q$-th neighbor. However, as first pointed out by Nataf \emph{et al.}~\cite{nataf2016exact}, these preliminary numerical evidences are not conclusive, and subsequent numerical validation only appears recently in SU(3) symmetry~\cite{gozel2020haldane, nataf2021edge}, where the energy gap in representation [3~0~0] and SU(3)$_1$ Wess-Zumino-Novikov-Witten (WZNW) university class in representation [2~0~0]  are confirmed. These results are consistent with field theory analysis by Affleck \emph{et al.}, which show that for the totally symmetric SU($N$) representation $[p~0~0]$, when $p$ is coprime with $N$, the system exhibits gapless excitations whose critical behavior is described by the SU($N$)$_1$ WZNW model, while when $ p/N $ is an integer, the system is gapped and possesses a unique ground state~\cite{lajko2017generalization, wamer2020generalization}.

The rich structure of higher SU($N$) symmetries can also support new topological features in the Haldane phase of spin chains. It has been shown that gapped quantum spin chains with PSU($N$) = SU($N$)/$\mathbb{Z}_N$ symmetry possess $N$ distinct SPT phases~\cite{duivenvoorden2012discriminating, duivenvoorden2013topological, roy2018chiral}. Among them, one is topologically trivial, while the remaining $N-1$ are nontrivial topological phases, which can be distinguished by different gapless edge degrees of freedom. For example, under local adjoint representations, the system exhibits different topological phases with topological $\mathbb{Z}_N$ quantum numbers $\pm 1$ or $\pm 2$, when the virtual spins transform under the fundamental or rank-2 antisymmetric representation of SU($N$) symmetry~\cite{roy2018chiral}.

To search for the above novel higher SU($N$) phases, the smallest value of $N$ one needs to take would be $N=3$. Under the SU(3) symmetry, as predicted by previous research, the simplest representations that support the Haldane gap are $[3~0~0]$ and $[2~1~0]$. For the former representation, large-scale DMRG simulations have identified a very small excitation gap of approximately $0.04J$~\cite{gozel2020haldane}. However, further studies based on field theory and variational matrix product state (MPS) methods suggest that the ground state actually corresponds to a trivial SPT phase~\cite{fromholz2020symmetry,devos2022haldane}. For the latter one, Monte Carlo simulations combined with field theory analyses suggest that the system is gapped and lies in a topologically nontrivial phase~\cite{wamer2019self}. This ground state is protected by the $ \mathbb{Z}_3 \times \mathbb{Z}_3 $ symmetry and shows a double degeneracy, with edge states in the representations $[1~0~0] - [1~1~0]$ or $[1~1~0] - [1~0~0]$, corresponding to inversion symmetry broken SPT states termed as the chiral Haldane phases~\cite{roy2018chiral, morimoto2014z, rachel2010spontaneous, wamer2019self}.

On the experimental side, alkaline-earth fermionic atoms offer a promising platform for realizing SU($N$) symmetry and above SPT phases~\cite{cazalilla2014ultracold}. Proposals are mainly focused on SPT states of various representations that are realized in the strong-coupling limit of lattice fermions~\cite{nonne2013symmetry, bois2015phase, fromholz2019haldane, madasu2025experimental}. Beyond the fermionic systems, the Haldane phase in the SU(3) adjoint representation could also be constructed using a two-species spinor Bose gas with species-dependent zigzag lattices~\cite{xu2023spinor}.

While these proposals provide various routes to realize the SU(3) spin chains, constructing corresponding SPT phases often relies on carefully engineered Hamiltonians. In this work, we propose a conceptually simpler and more natural setting of the chiral Haldane phases, by constructing the SU(3) AFH model with alternating conjugate representations (fundamental and anti-fundamental) through mapping to the Holstein-Primakoff bosons.

 The alternating conjugate representation in SU(3) symmetry displays nontrivial low-energy behavior~\cite{affleck1985large, affleck1990exact}, in contrast to the SU(2) case, where the equivalence of fundamental and anti-fundamental representations leads to a quantum critical phase described by the SU(2)$_1$ WZNW theory. We will first construct the SU(3) AFH spin chain with alternating fundamental and anti-fundamental representations by means of a linear transformation. The presence of a finite excitation gap and exponentially decaying correlation functions is subsequently confirmed via numerical calculations. By introducing staggered interactions on even and odd bonds, we demonstrate that the system is connected to the chiral Haldane phases with opposite chiralities. These two topological phases are distinguished by string orders that are defined in different unit cells. Specifically, we observe that there exists a narrow region near the Heisenberg point, where not only the ground state but also the first excited state exhibit nontrivial topological properties. Unlike systems with strong disorder~\cite{basko2006problem, pal2010many, chandran2014many, potter2015protection}, we find that the topological nature of this first excited state here originates from the level crossing between two competing SPT states. This provides a low-energy manifestation of SPT order that remains stable without the need for disorder.

This paper is organized as follows.  
In Sec.~\ref{bosemap}, we introduce a bosonic realization of the SU(3) AFH spin chains with local alternating conjugate representation by Holstein-Primakoff bosons.  
In Sec.~\ref{phasediagram}, we include a staggered interaction strength and anisotropy along the $T^3$ and $T^8$ directions in the Hamiltonian, and give the quantum phase diagram of the system where we observe two topological nontrivial phases with opposite chiralities and a $\mathbb{Z}_3$ symmetry-breaking phase. 
Section~\ref{topotransition} is devoted to the chiral-reversed phase transition. We confirm that it is a first-order quantum phase transition, and observe an excited-state SPT phase around the Heisenberg point.
In Sec.~\ref{symmetrybreaking}, we consider the quantum phase transition associated with the spontaneous symmetry breaking of $\mathbb{Z}_3$ symmetry. We give a variational wavefunction ansatz, and show that it provides a consistent description of this phase transition.
In Sec.~\ref{experiment}, we give an experimental proposal to realize our model by making use of spin-1/2 bosons.

\section{The bosonic mapping}\label{bosemap}
We first consider the following SU(3) alternating conjugate representation AFH chain~\cite{auerbach2012interacting}

\begin{align}
	H_S= J \sum_{i=1}^{L/2} \sum_{\delta=\pm1} \mathcal{S}_{2i}^{\alpha\beta} \, \overline{\mathcal{S}}_{2i+\delta}^{\beta\alpha}
	\label{eq:lieham}
\end{align}
with $J>0$, and $\delta=\pm 1$ labels the even or odd bonds for later convenience. Here the summations over repeated indices ($\alpha, \beta = 1, 2, 3$) are assumed, and $S^{\alpha\beta}$ ($\overline{S}^{\alpha\beta}$) are the SU(3) generators with fundamental (anti-fundamental) representation which obey the following commutation relations
\begin{align}
	\left[\mathcal{S}^{\alpha\beta},\mathcal{S}^{\beta'\alpha'}\right]= \delta_{\beta\beta'}\mathcal{S}^{\alpha\alpha'}-\delta_{\alpha'\alpha}\mathcal{S}^{\beta'\beta}.
\end{align}
It is more convenient to rewrite these generators in terms of an eight-component spin operator $\bm{S}$ with elements $S^a$ ($ a = 1, \dots, 8 $) as
\begin{align}
	\mathcal{S}^{\alpha\beta}=\sum_{a=1}^8\left(\lambda^a\right)_{\beta\alpha}S^a,
\end{align}
where $ \lambda^{a} $ are the Gell-Mann matrices. It is now straightforward to see that the Hamiltonian Eq.~(\ref{eq:lieham}) is equivalent to
\begin{align}\label{eq:spinham}
	H_S = 2J\sum_{i=1}^{L/2} \sum_{\delta = \pm 1} \bm{S}_{2i} \cdot \overline{\bm{S}}_{2i + \delta}
\end{align}
with $\bm{S}$ and $\overline{\bm{S}}$ the spin operator corresponding to fundamental and anti-fundamental representations on even and odd sites, respectively, and
\begin{align}
	 S_{i}^{a} = \frac{1}{2} \lambda_{i}^{a}, \quad   \overline{S}_{i}^{a} = \frac{1}{2} \overline{\lambda}_{i}^{a},
\end{align}
where we have $\overline{\lambda} = -\lambda^{*}$. To see the connection to Holstein-Primakoff bosons, it is more convenient to introduce the ladder operators $ T^{\pm} = S^{1} \pm i S^{2} $, $ V^{\pm} = S^{4} \pm i S^{5} $, and $ U^{\pm} = S^{6} \pm i S^{7} $, after which the above Hamiltonian can be written as
\begin{align}\label{eq:spinhamsym}
	H_S =&J\sum_{i=1}^{L/2} \sum_{\delta = \pm 1} \left[ \left(T_{2i}^{+} \overline{T}_{2i+\delta}^{-} + U_{2i}^{+} \overline{U}_{2i+\delta}^{-} + V_{2i}^{+} \overline{V}_{2i+\delta}^{-} \right.\right. \nonumber \\
	&\left.\left. + \text{H.c.}\right) + 2T_{2i}^{3} \overline{T}_{2i+\delta}^{3} + 2T_{2i}^{8} \overline{T}_{2i+\delta}^{8} \right],
\end{align}
where operators $T^{3} = S^{3}$ and $T^{8} = S^{8}$ provide two independent conserved quantum numbers of this system.

\begin{figure}[t]
	\centering
	\includegraphics[width=1\linewidth]{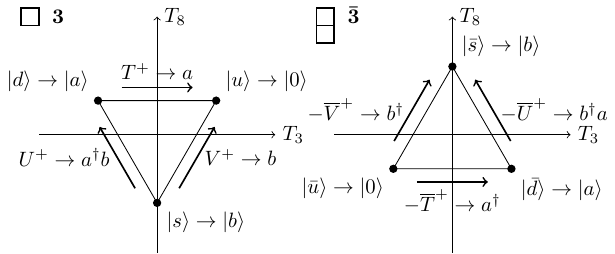}
	\caption{The weight diagram of SU(3) triplet state in the Holstein-Primakoff bosons representation. The Hilbert space is spanned by two bosonic states $\vert a\rangle$ and $\vert b\rangle$, together with a vacuum state $\vert 0\rangle$, corresponding to the $|u\rangle, |d\rangle, |s\rangle$ states in the fundamental representation $\bm{3}$, or $|\overline{u}\rangle, |\overline{d}\rangle, |\overline{s}\rangle$ states in the anti-fundamental representation $\overline{\bm{3}}$. The rising and lowering operators can be constructed by the creation or annihilation of bosonic particles.}
	\label{fig:boserep}
\end{figure}

The Holstein-Primakoff transformation is carried out as follows. We construct the local Hilbert space (either with SU(3) fundamental or anti-fundamental representation) using a bosonic system with two kinds of hardcore bosons ($a$ and $b$), subject to the single-occupancy constraint $n_a + n_b \leq 1$. In this way, we get a one-to-one correspondence between the SU(3) triplet states to bosonic spaces $\vert0\rangle$, $\vert a\rangle$, and $\vert b\rangle$ (see Fig.~\ref{fig:boserep}). The SU(3) ladder operators can then be constructed by the bosonic creation and annihilation operators, as illustrated in Fig.~\ref{fig:boserep}. For example, if we label $\vert 0\rangle$ and $\vert a\rangle$ states as the $\vert u\rangle$ and $\vert d\rangle$ states in the SU(3) fundamental representation $\mathbf{3}$, the ladder operator $T^+$ can be represented by annihilating a bosonic particle $a$. Since in the anti-fundamental representation $\overline{T}^+=-T^-=-a^\dagger$, the states $\vert\overline{u}\rangle$ and $\vert\overline{d}\rangle$ would then correspond to bosonic states $\vert 0\rangle$ and $\vert a\rangle$.

Based on the above arguments, we prove the following mapping from the SU(3) fundamental representation to the Holstein-Primakoff bosons
\begin{align}
	&T^{+} =  a, \quad T^{-} = a^{\dagger}, \nonumber\\
	&V^{+} =  b, \quad V^{-} = b^{\dagger}, \nonumber\\
	&U^{+} = a^{\dagger} b, \quad U^{-} = b^{\dagger} a ,\nonumber\\
	&T^{3} = \frac{1}{2}(1 - b^{\dagger} b) - a^{\dagger} a, \nonumber\\
	&T^{8} = \frac{\sqrt{3}}{2} \left( \frac{1}{3} - b^{\dagger} b \right) .
\end{align}
We note that the above mapping is a linear transformation, and it is exact only in SU(3) triplet representation. For higher-dimensional representations, the mapping can only be realized in a non-linear form as provided in~\cite{wagner1975nonlinear}. Substituting above transformation into our Hamiltonian Eq.~(\ref{eq:spinham}), we get the bosonic realization of the SU(3) AFH chain as
\begin{align}
	H_S =&J \sum_{i=1}^{L} \Bigg[ 
		-\frac{2}{3} + \left(a_{i}^{\dagger} a_{i}+ a_{i+1}^{\dagger} a_{i+1}+ b_{i}^{\dagger} b_{i}+ b_{i+1}^{\dagger} b_{i+1}\right) \nonumber\\
	& - 2\left(a_{i}^{\dagger} a_{i} a_{i+1}^{\dagger} a_{i+1}+ b_{i}^{\dagger} b_{i} b_{i+1}^{\dagger} b_{i+1}\right) - \left(a_{i}^{\dagger} a_{i} b_{i+1}^{\dagger} b_{i+1}\right. \nonumber\\
	&\left.+ b_{i}^{\dagger} b_{i} a_{i+1}^{\dagger} a_{i+1}\right)-\left(a_{i} a_{i+1}+ b_{i} b_{i+1} +\text{H.c.}\right) \nonumber\\
		& -\left(a_{i}^{\dagger} a_{i+1}^{\dagger} b_{i} b_{i+1}+\text{H.c.}\right) 
	\Bigg].
	\label{eq:projham}
\end{align}

To show that our Hamiltonian Eq.~(\ref{eq:projham}) preserves the $\mathbb{Z}_3 \times \mathbb{Z}_3$ symmetry, we define operators for the cyclic permutation of quark states $P$ and the phase-rotation $Q$ as
\begin{align}
	P &= |0\rangle\langle a|+|a\rangle\langle b|+|b\rangle\langle 0|=a+a^\dagger b+b^\dagger,\\		
	Q &= |0\rangle \langle 0|+e^{2\pi i/3} |a\rangle\langle a|+e^{4\pi i/3}|b\rangle\langle b|\nonumber\\
	   &=\exp\left[\frac{2\pi i}{3} \left(n_a + 2 n_b\right)\right],
\end{align}
from which we have
\begin{align}
	P^3 = Q^3 = \mathbb{I},\quad PQ=e^{2\pi i/3} QP.
\end{align}	
For the anti-fundamental representation, we have $\overline{P} = P$, and $\overline{Q} = Q^{*}$, and it's straightforward to show that $\left[H_S,P+\overline{P}\right]=\left[H_S,Q+\overline{Q}\right]=0$.

\section{The Phase Diagram}\label{phasediagram}
To see the connection with different chiral Haldane phases and explore their quantum phase transition, in this work we consider a more general Hamiltonian 
\begin{align}\label{eq:fullham}
	H = J_{R} \sum_{i=1}^{L/2} H_{2i, 2i+1}+J_{L} \sum_{i=1}^{L/2} H_{2i, 2i-1} 
\end{align}
with
\begin{align}
	H_{i,j} =& -\frac{2g}{3} + g\left(a_{i}^{\dagger} a_{i} + a_{j}^{\dagger} a_{j} + b_{i}^{\dagger} b_{i} + b_{j}^{\dagger} b_{j}\right) \nonumber\\
	& - 2g\left(a_{i}^{\dagger} a_{i} a_{j}^{\dagger} a_{j} + b_{i}^{\dagger} b_{i} b_{j}^{\dagger} b_{j}\right) - g\left(a_{i}^{\dagger} a_{i} b_{j}^{\dagger} b_{j} + b_{i}^{\dagger} b_{i} a_{j}^{\dagger} a_{j}\right) \nonumber\\
	& - \left(a_{i} a_{j} + b_{i} b_{j} + \text{H.c.}\right) 
		- \left(a_{i}^{\dagger} a_{j}^{\dagger} b_{i} b_{j} + \text{H.c.}\right),
\end{align}
where we have included a staggered interaction $J_R$ and $J_L$ on even and odd bonds. The letter $R$ or $L$ corresponds to the case when $J_R$ or $J_L$ are dominated, the system is connected to a right or left-chiral Haldane phase. We also include an anisotropic coupling strength $g$ in $T^3$ and $T^8$ terms as appeared in Eq.~(\ref{eq:spinhamsym}). Similar to the spin XXZ model, this term breaks the SU(3) symmetry, and drives the system away from the topological Haldane phases and into a trivial phase with spontaneous $\mathbb{Z}_3$ symmetry breaking.

To give the phase diagram of Hamiltonian Eq.~(\ref{eq:fullham}), we define the following string order parameter for the right and left-chiral Haldane phases as
\begin{align}\label{eq:string}
	{\cal O}^{\rm str}_{R/L} (k-j)=\left\langle {\cal O}^{R/L}_j \exp \left( i\pi \sum_{j<l<k} {\cal O}^{R/L}_l \right) {\cal O}^{R/L}_k \right\rangle,
\end{align}
where the operators ${\cal O}^{R/L}_i={\cal O}^0_{2i} - {\cal O}^0_{2i\mp 1}$ are defined on different two-site unit cells with
\begin{equation}
	{\cal O}^0_i=\frac{2}{3}-n^a_i-n^b_i.
	\label{eq:unitop}
\end{equation}
We note that there have been various definitions of the string order in higher SU(N) symmetries~\cite{duivenvoorden2012discriminating, duivenvoorden2013symmetry}, and our definition here is equivalent to the one given in~\cite{morimoto2014z}.

Our definition of the string order can be easily understood from the following limit. Considering $J_R/J_L\to 0$ with $g=1$, the system becomes a fully dimerized phase with each odd bond forming a SU(3) singlet, and this ground state can be written as
\begin{align}
	|\Psi\rangle_L=\left[ \frac{1}{\sqrt{3}} \left( |0 0 \rangle +  |a a \rangle +  |b b \rangle \right) \right]^{\otimes \frac{L}{2}},
\end{align}
which is just the left-chiral AKLT state on the even-bond unit cells, and supports a left-chiral string order $\left| {\cal O}^{\rm str}_L \right| = 16/81$. The opposite limit occurs at $J_R/J_L\to \infty$, and the system is described by a right-chiral AKLT state, with an odd-bond unit cell. Thus, at these two limits, the system is connected to left and right-chiral Haldane phases, which are described by our definition of string order Eq.~(\ref{eq:string}).

\begin{figure}[htb]
	\centering
	\includegraphics[width=1.0\linewidth]{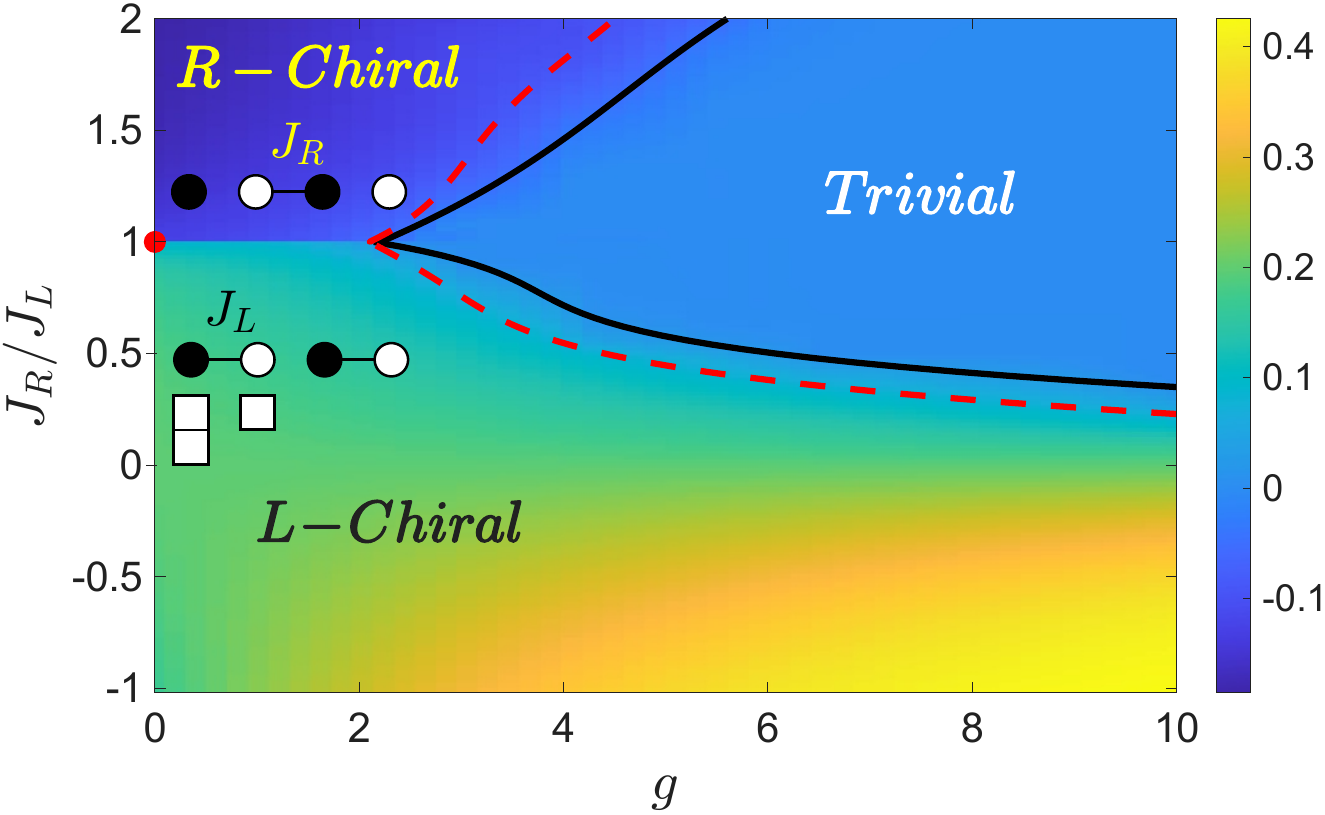}
	\caption{The phase diagram as a function of staggered interaction $J_R/J_L$ and anisotropy $g$. The color scale indicates the magnitude of the string order parameter with system size $L = 180$ and bond dimension $m=600$ under open boundary conditions. From the string order, we observe two SPT states with different chiralities and a trivial phase. There is a chiral-reversed transition at $J_R/J_L=1$, and the solid curve denotes the phase boundary between two SPT phases to the trivial phase that is determined by the change of average particle number. This transition can be qualitatively explained by our variational ansatz (the red dashed lines) as detailed in Sec.~\ref{ansatz}.}
	\label{fig:phasediagram}
\end{figure}

To find the ground state of our system, we carry out DMRG calculations based on the Algorithms and Libraries for Physics Simulations (ALPS) libraries~\cite{albuquerque2007alps, bauer2011alps, dolfi2014matrix}. Figure~\ref{fig:phasediagram} shows our phase diagram of Hamiltonian Eq.~(\ref{eq:fullham}), where the colored density plot labels the magnitude of the string order parameter. To distinguish between the two different chiralities, we define the left-chiral string order to be positive, while the right-chiral one to be negative.
We observe a chiral-reversed transition at $J_R/J_L=1$ for small interaction anisotropy $g$, and a topological to trivial phase transition for sufficiently strong $g$. The latter transition can also be determined by the average particle number (solid line), through which the average particle number changes abruptly from $1/3$ for $a$ or $b$ bosons to full occupation of either type of bosons or in a vacuum state, which will be explained in detail later.

To see the connection to the chiral Haldane phase in the SU(3) adjoint representation AFH chain, we extend the phase diagram to the negative $J_R/J_L$ regime. As expected, the chiral Haldane phase can be adiabatically extended to the negative $J_R/J_L$ regime. In the limit $J_R/J_L \to -\infty$, each pair of $ \bm{3} $ and $ \overline{\bm{3}} $ representations that coupled by $J_R$ forms an octet state, and the system reduces to an SU(3) AFH spin chain in the adjoint representation, with the string order parameter reaching the value $ \left| {\cal O}^{\rm str}_{0} \right| = 1/4 $~\cite{morimoto2014z}.

We pay special attention to the transition point $J_R/J_L=1$ and $g=0$ (shown as the red point in Fig.~\ref{fig:phasediagram}). Since at the Heisenberg point ($J_R/J_L=1$, $g=1$), previous studies have confirmed a rather small energy gap ~\cite{affleck1985large, affleck1990exact}, it would be nontrivial to ask whether this small gap still survives extending to $g=0$.

\begin{figure*}[htbp]
	\centering
	\includegraphics[width=1\textwidth]{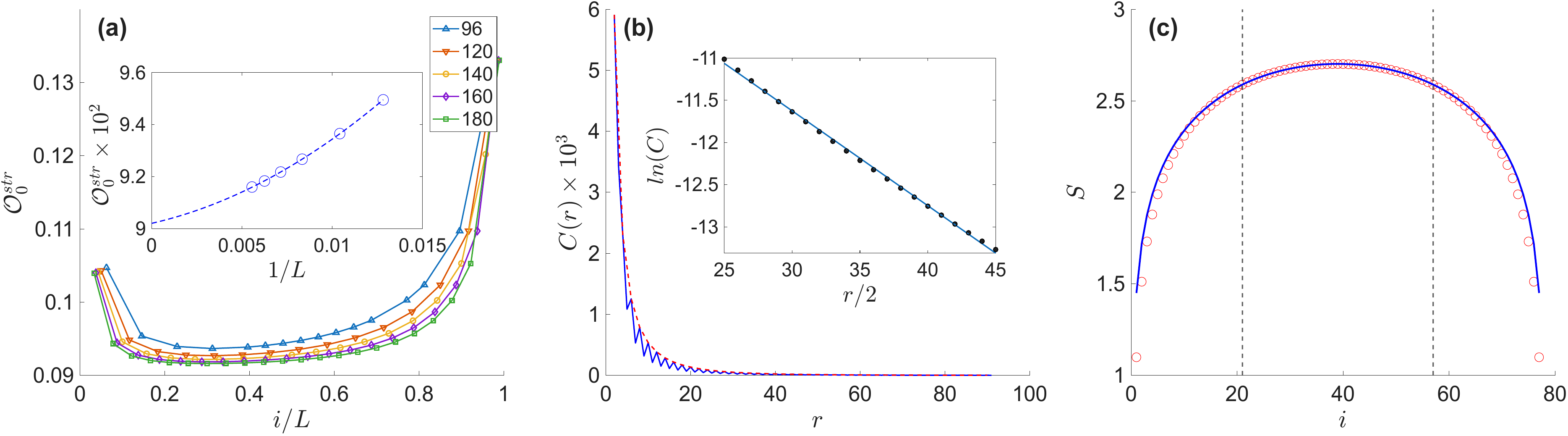}
	\caption{(a) The string order parameter for different system sizes at $J_R/J_L = 1 $ and $ g = 0 $ (OBC, $ m = 800 $). The inset shows a second-order polynomial fitting as a function of $1/L$, which gives a finite string order of $\mathcal{O}_{0}^{\text{str}} = 0.090 \pm 0.001 $ at the thermodynamic limit. (b) The correlation function as a function of distance ($L=180$, OBC, $m=800$). The correlation function at odd data points (dashed red line) shows a well-behaved exponential decay, and the exponential fitting in the inset gives a correlation length of approximately 15.8. (c) The subsystem entanglement entropy for $J_R / J_L = 1 $ and $ g = 0 $ (PBC, $ L = 78 $, $ m = 1400 $). The blue solid line corresponds to our fitting using Eq.~(\ref{eq:entropy}), performed around the chain center (see the vertical dashed lines), which gives a central charge ($c \approx 1.17$) that is below the value predicted by SU(3)$_1$ WZNW critical theory.}
	\label{fig:corr}
\end{figure*}

 We show our results at this point in Fig.~\ref{fig:corr}, confirming that the ground state is away from critical. In Fig.~\ref{fig:corr}a we show our results of the string order parameters for various system sizes, and by interpolating to the thermodynamic limit we find a finite string order $\mathcal{O}_{0}^{\text{str}} = 0.090 \pm 0.001 $, indicating that the system still shows topological behaviors that are protected by the Haldane gap. In Fig.~\ref{fig:corr}b, we observe an exponentially decaying correlation function at long distances, which is consistent with a gapped phase.

 We also calculate the entanglement entropy of the system in Fig.~\ref{fig:corr}c. We do not observe a clear plateau structure as expected for a gapped phase for the system size and bond dimension we considered. However, by fitting to the entanglement entropy of a critical system using~\cite{calabrese2004entanglement}
\begin{equation}\label{eq:entropy}
	S_{L} (\alpha)= \frac{c}{3} \log \left[ \left( \frac{L}{\pi} \right) \sin \left( \frac{\pi \alpha}{L} \right) \right] + c_1,
\end{equation}
we obtain a central charge of approximately $ c \approx 1.17$. According to the relation $ c = k(n^2 - 1)/(k + n) $, we can find that this central charge is already significantly below the theoretical lower bound of $c=2$ described by the critical theory of SU(3)$_{k=1}$ WZNW model, which also indicates the system is away from the critical phase and supports a finite energy gap.

\section{Chiral Topological transition}\label{topotransition}
In this section, we consider the transition between different chiral topological phases. We calculate the energy gap along this transition line and confirm that this transition belongs to the first order. We then show that there exist excited-state topological phases that are protected by $\mathbb{Z}_3 \times \mathbb{Z}_3$ symmetry.

\subsection{The chiral-reversed transition}
\begin{figure}[t]
	\centering
	\includegraphics[width=0.85\linewidth]{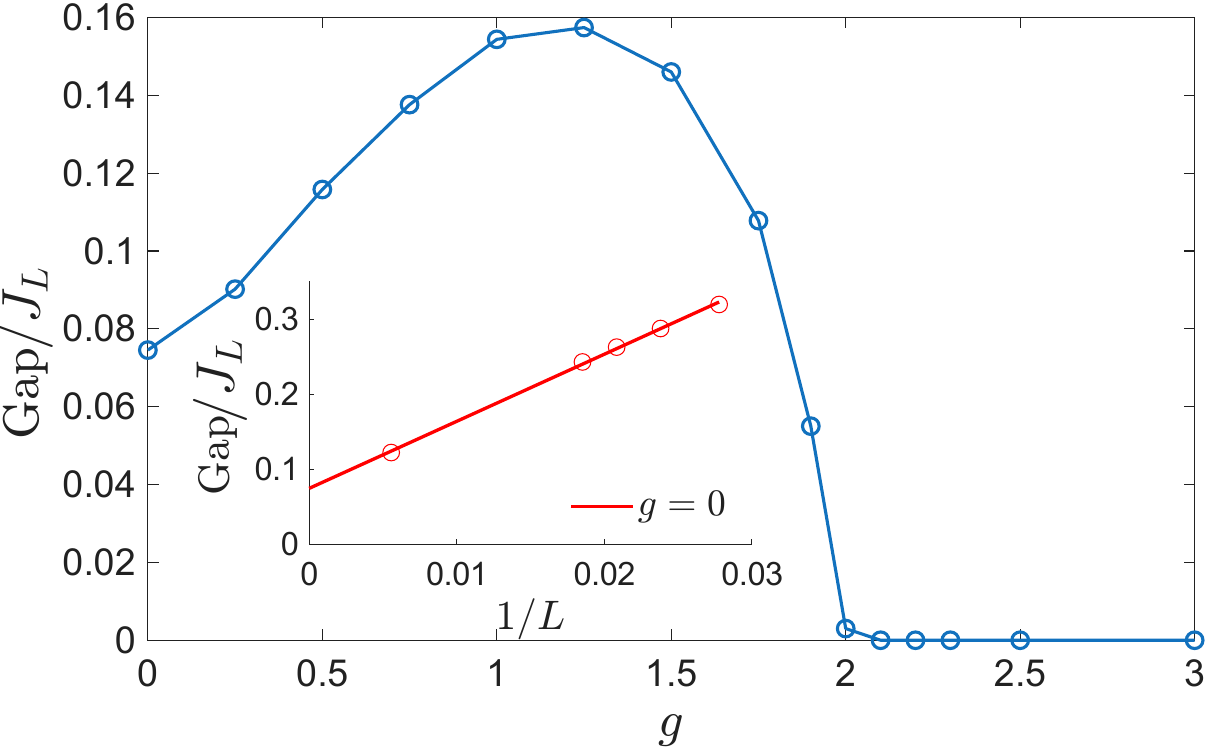}
	\caption{The energy gap as a function of $ g $ at $J_R/J_L = 1$ (OBC, $ m = 800 $). The data points show the results by polynomial fitting the gap to the thermodynamic limit, as illustrated in the inset. At the point $ g = 0 $, we find an energy gap of approximately $0.066 \pm 0.001 $.}
	\label{fig:gap}
\end{figure}

We calculate the energy gap along the chiral transition line as illustrated in Fig.~\ref{fig:gap}. The energy gap extends to around $g=2$, consistent with our phase diagram Fig.~\ref{fig:phasediagram}. As the results in Fig.~\ref{fig:corr} have shown, the transition point at $g=0$ is away from critical, here we fit the energy gap at this point and get approximately $0.066$. The reason for this gapped chiral transition line from $g=0$ to $g\approx 2$ is due to its first-order nature, and we confirm this in the following.

\begin{figure}[h]
	\centering
	\includegraphics[width=0.8\linewidth]{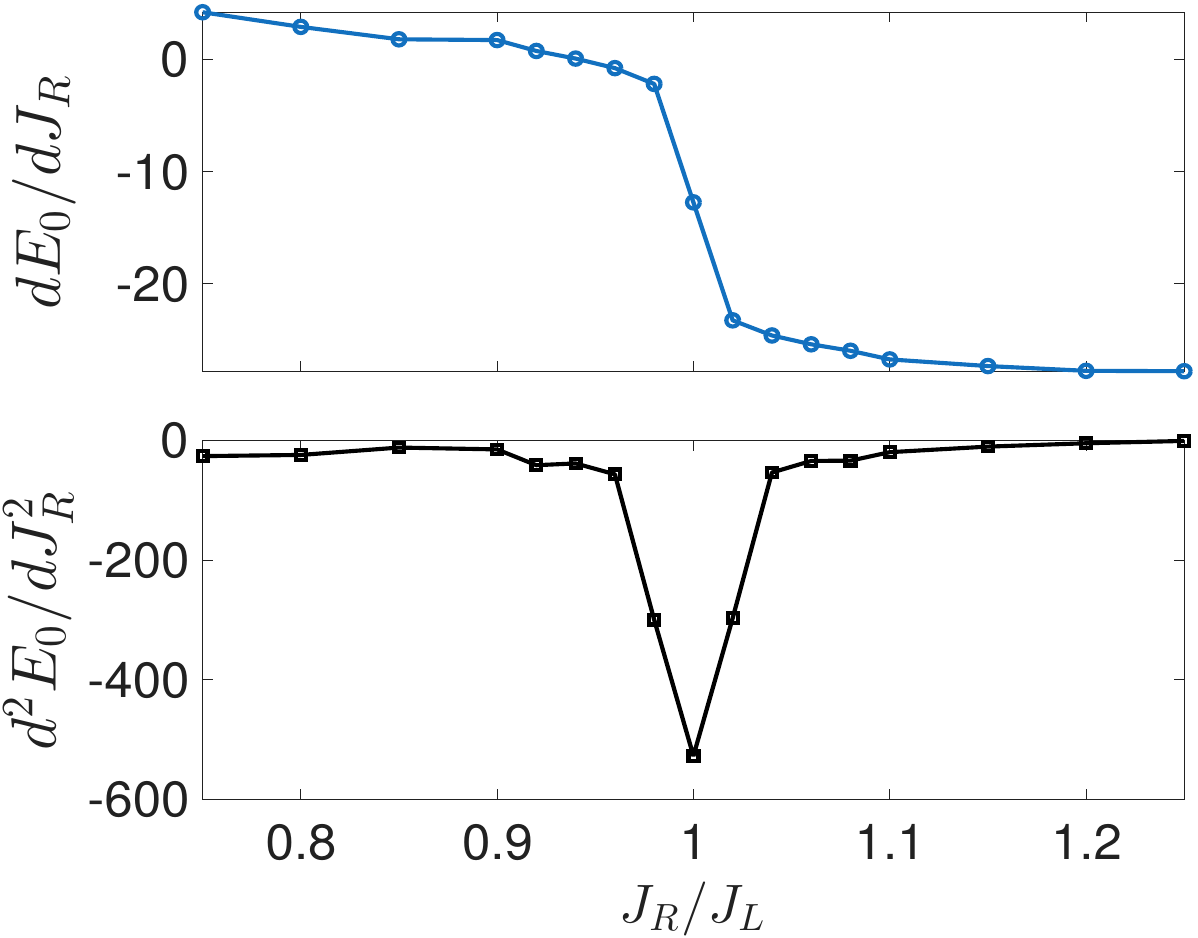}
	\caption{The first and second-order derivatives of the ground-state energy $E_0$ to $J_R$ as a function of $J_R/J_L$ at $g=1$ (PBC, $L=64$, $m=800$). The discontinuities of both derivatives at $J_R/J_L=1$ signal a first-order quantum phase transition.}
	\label{fig:derivative}
\end{figure}

We calculate the first and second-order derivatives of the ground-state energy $E_0$ to $J_R$ across this chiral transition at $g=1$, as illustrated in Fig.~\ref{fig:derivative}. As $J_R/J_L$ increases to the transition point $J_R/J_L=1$, we observe a discontinuity in the first derivative of $E_0$. This discontinuity leads to a delta function at the transition point in the second derivative. These results provide evidence that the gapped chiral transition here belongs to a first-order quantum phase transition. We note that since the $\mathbb{Z}_3 \times \mathbb{Z}_3$ symmetry is preserved all the way throughout the chiral transition, the ground state of the system then remains symmetry-protected across this transition.

\subsection{SPT in the first excited state}
\begin{figure}[hb]
	\centering
	\includegraphics[width=0.8\linewidth]{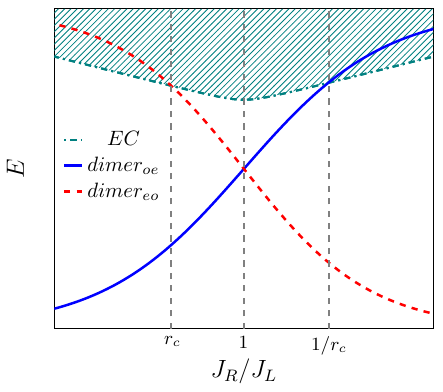}
	\caption{Schematic energy structure of the system under PBC, showing the evolution of energy levels as a function of $ J_R/J_L $ at $g=1$. The red dashed line and the blue solid line represent two distinct topological states with different chiralities. The shaded region above the green dash-dotted line represents a higher excitation continuum. At the transition point $ J_R/J_L = 1 $, the system exhibits a twofold degeneracy. Approaching the transition point, the two chiral states become the ground and first excited states, respectively, and they are both protected by a finite energy gap. At $ L = 48 $, we obtain $r_c = 0.980 \pm 0.001 $. Due to numerical limitations, we are unable to locate the position of $r_c$ for larger $L$, however, given the existence of a finite energy gap, we believe the position should be very close to but always away from the transition point. }
	\label{fig:schematic}
\end{figure}

SPT phases are typically defined in the ground states where the excitations are blocked. However, recent researches show that in systems with many-body localization (MBL), symmetry-protected topological behaviors can also appear in high energy levels~\cite{chandran2014many, potter2015protection}. In this paper, we propose another mechanism for the excited-state SPT states, that is driven by the first-order quantum phase transition between two distinct topological phases.

We illustrate the schematic energy structure of our system in Fig.~\ref{fig:schematic}. At the Heisenberg point ($J_R = J_L, g=1$), the system shows a double degeneracy of two chiral Haldane phases (blue solid and red dashed lines), which is gapped from the higher excited continuum (green shaded region), as illustrated in Fig.~\ref{fig:schematic}. Slightly away from this point ($r_c<J_R/J_L<1/r_c$), these two chiral phases become ground and first excited states respectively, while still gapped from the higher excited continuum. In this region, not only the ground state, but also the first excited states are symmetry-protected. Outside this region, the first excited state merges into the higher excited continuum, and thus loses its topological protection.

\begin{figure}[t]
	\centering
	\includegraphics[width=0.85\linewidth]{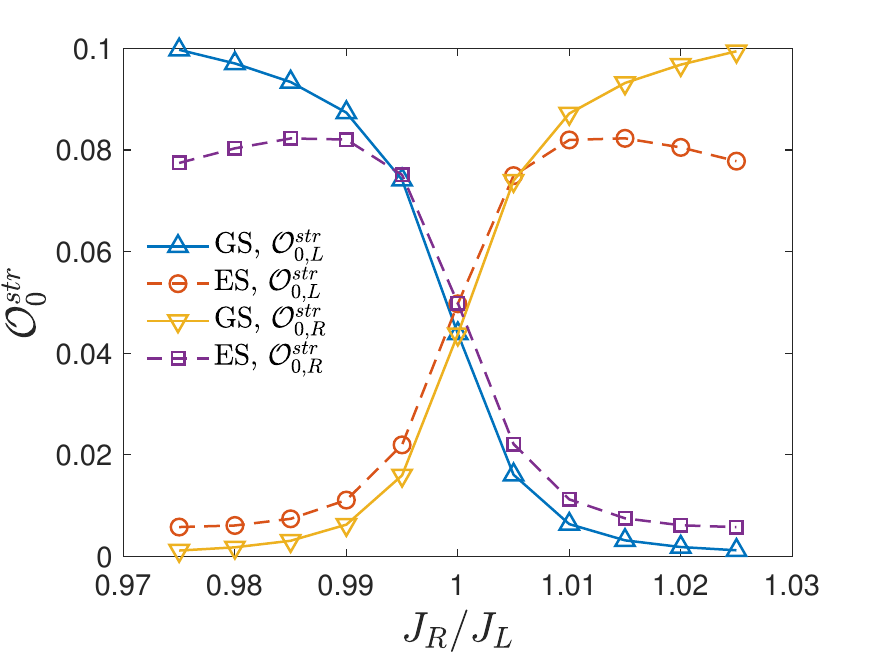}
	\caption{The string order parameters for both the ground state and the first excited state as a function of $ J_R/J_L $ at $g=1$ (PBC, $ L = 36 $, $ m = 500 $). Approaching the transition point with $J_R/J_L=1$, both the ground (solid line) and first excited (dashed line) states with opposite chirality show a finite string order, while deviating from this point the string order in the first excited state decreases and will be destroyed by the continuum, consistent with our qualitative energy structure in Fig.~\ref{fig:schematic}.}
	\label{fig:excitedorder}
\end{figure}

To show that the first excited state in this region also exhibits nontrivial topological properties. We plot the corresponding string orders with opposite chirality in Fig.~\ref{fig:excitedorder}. We can find that both the ground and first excited states support finite string order around the chiral transition point $J_R/J_L=1$, meaning that they all hold nontrivial topological behaviors. Further away from this transition point, we observe the string order in the first excited state tends to decrease, which is consistent with the above argument that this first excited state will eventually merge into the higher excited continuum, and lose its topological behavior.

\begin{figure}[t]
	\centering
	\includegraphics[width=0.85\linewidth]{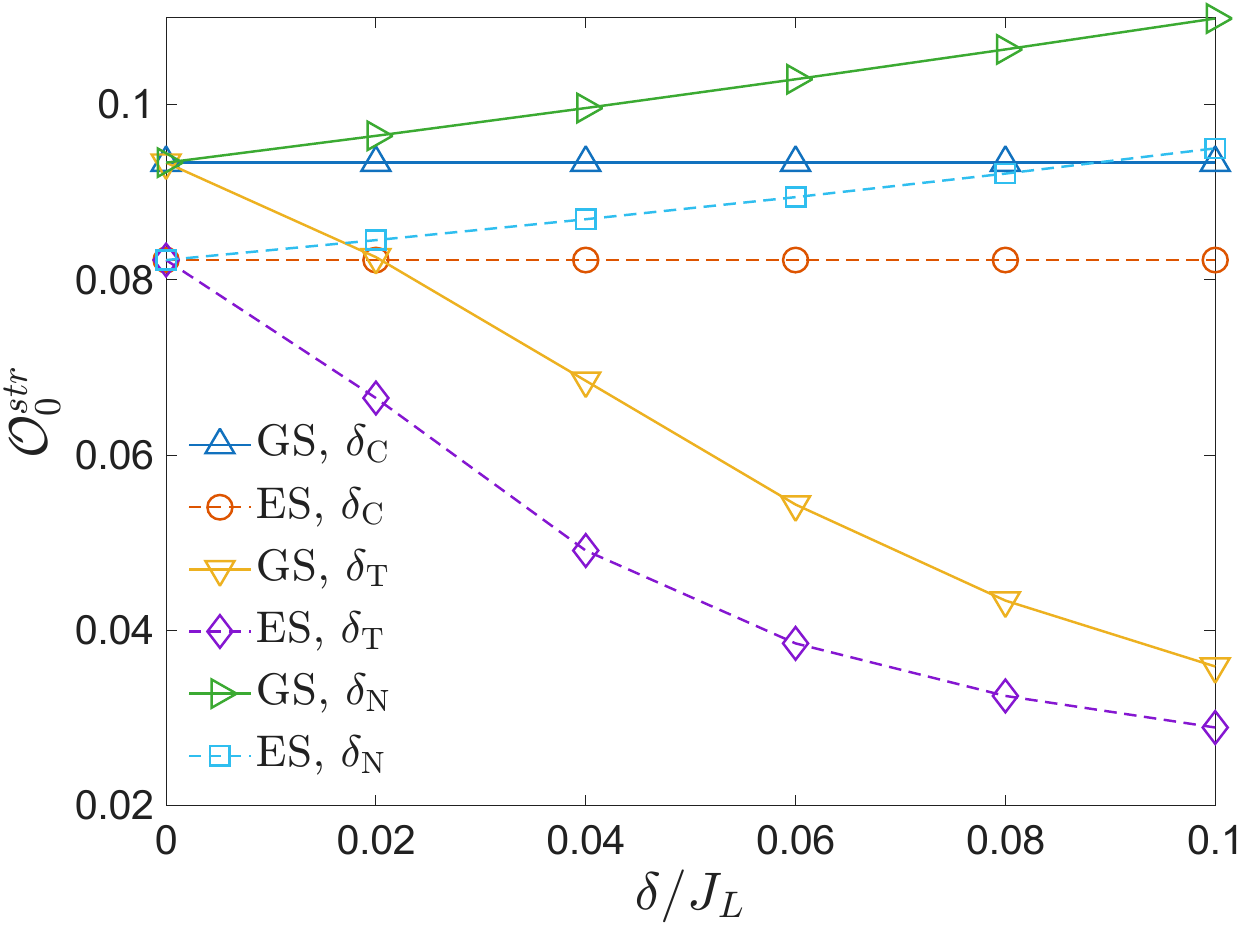}
	\caption{Symmetry protection of the ground and first excited states around the transition point with $g=1$ and $J_R/J_L=0.985$ (PBC, $ L = 36 $, $ m = 500 $). We show the string orders of both states under three different perturbations $\delta_C$, $\delta_N$, and $\delta_T$ that preserve or break the $\mathbb{Z}_3 \times \mathbb{Z}_3$ symmetry, respectively. The string orders remain robust under the symmetry-preserved perturbation $\delta_C$ ($\delta_N$), while they decay rapidly under a small perturbation $\delta_T$, indicating that both states are SPT phases and are protected by the $\mathbb{Z}_3 \times \mathbb{Z}_3$ symmetry.}
	\label{fig:perturbation}
\end{figure}

To prove that this first excited state is symmetry protected, we consider two types of perturbations, one conserves the $\mathbb{Z}_3 \times \mathbb{Z}_3$ symmetry with $H'_{\mathcal{C}}=\delta_{\mathcal{C}}\sum_i\mathcal{C}_{2,i}$ or $H'_{N}=\delta_N\sum_i \bm{S}_{i}\bm{S}_{i+2} $, the other breaks one of the  $\mathbb{Z}_3$ symmetry with $H'_{T}=\delta_T\sum_iT^8_i$, where
\begin{align}
	\mathcal{C}_{2,i}=&\frac{1}{2} \left(a_i^\dagger a_i + a_i a_i^\dagger + b_i^\dagger b_i + b_i b_i^\dagger + a_i^\dagger b_i b_i^\dagger a_i + b_i^\dagger a_i a_i^\dagger b_i\right)\nonumber\\
	&+ \left(n^{a}_{i}\right)^2 + \left(n^{b}_{i}\right)^2 + n^{a}_{i} n^{b}_{i} - n^{a}_{i} - n^{b}_{i}
	\label{eq:casimir}
\end{align}
corresponds to the local Casimir operator,
\begin{align}
	\bm{S}_{i}\bm{S}_{i+2} =& \overline{\bm{S}}_{i}\overline{\bm{S}}_{i+2}\nonumber\\
	=&\frac{1}{3} -\frac{1}{2}\left( n^{a}_{i} + n^{a}_{i+2} + n^{b}_{i} + n^{b}_{i+2} \right)\nonumber\\
	 &+ n^{a}_{i}n^{a}_{i+2} + n^{b}_{i}n^{b}_{i+2} + \frac{1}{2}\left(n^{a}_{i} n^{b}_{i+2} + n^{b}_{i} n^{a}_{i+2}\right)\nonumber\\ 
	&+\frac{1}{2} \left(a_i^\dagger a_{i+2} + b_i^\dagger b_{i+2} + a_i^\dagger b_i b_{i+2}^\dagger a_{i+2} + \text{H.c.}\right)
\end{align}
accounts for the next-nearest-neighbor interaction, and
\begin{align}
	T^8_i=&\frac{1}{2}(1 - n^{b}_{i}) - n^{a}_{i}.
\end{align}
Note that we have kept track of the operators in the first terms of Eq.~(\ref{eq:casimir}) due to the non-trivial commutation relationship, and this terms serve as a check of our local representations.
We show the string order as a function of the strength of these three perturbations in Fig.~\ref{fig:perturbation}. As we can see, both the ground and first excited states show robust string orders corresponding to the perturbation $\delta_C$ ($\delta_N$), while decaying significantly when the perturbation $\delta_T$ is included, signaling the breaking of a topological state.

\section{Spontaneous $\mathbb{Z}_3$ symmetry breaking}\label{symmetrybreaking}
By increasing the interaction anisotropy $g$, we observe a spontaneous breaking of $\mathbb{Z}_3$ symmetry to a trivial phase. In this section, we first identify this quantum phase transition through its average particle number, then we provide a variational ansatz and give a qualitative explanation of this transition.

\subsection{Average particle number}
\begin{figure}[t]
	\centering
	\includegraphics[width=0.8\linewidth]{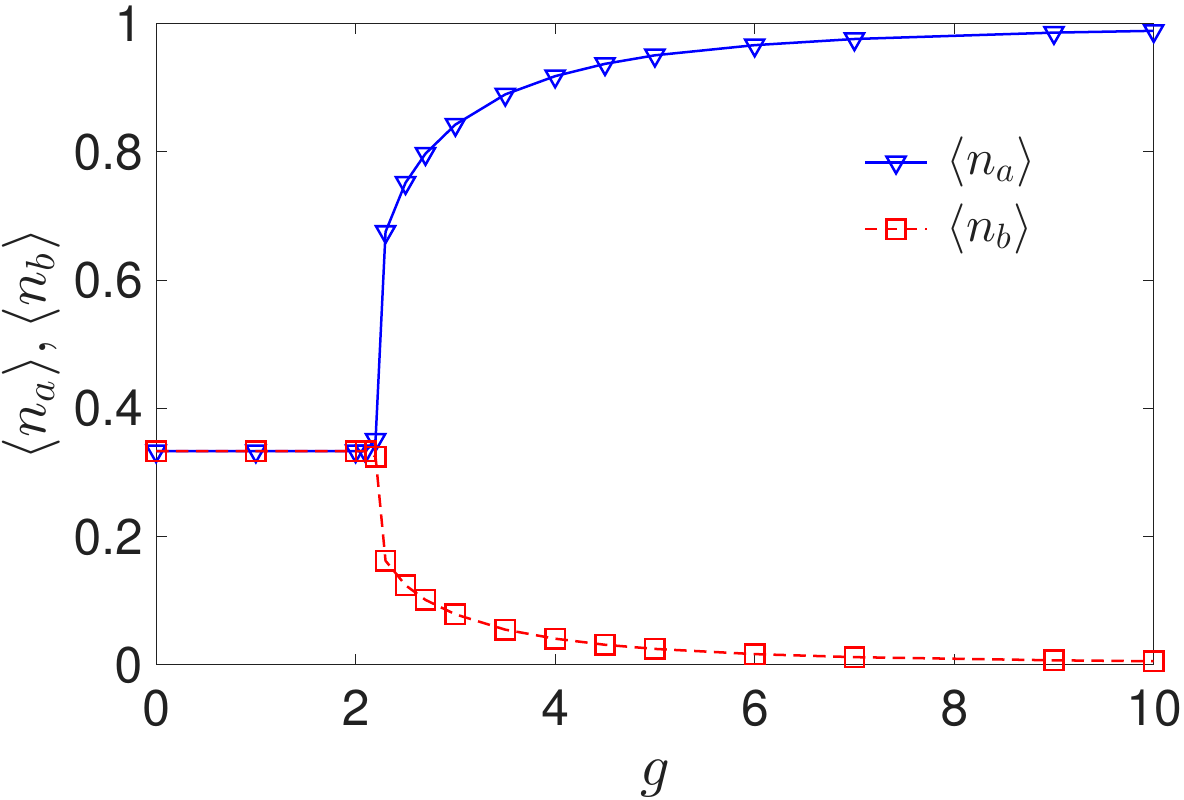}
	\caption{The average number of $a$ and $b$ particles per site as a function of $ g $ at $ J_R/J_L = 1 $ (OBC, $L=180$, $m=800$). As $g$ increases, we see a second-order phase transition around $g = 2.2 $, where the $\mathbb{Z}_3$ symmetry is spontaneously broken, and the system goes into one of its $\mathbb{Z}_3$ symmetry broken phases, either in $|a a \cdots \rangle$, $ |b b \cdots \rangle $, or $ |0 0 \cdots \rangle$ (Note in this figure we only show one of the resulting phases).}
	\label{fig:number}
\end{figure}

In Fig.~\ref{fig:phasediagram}, we observe a second-order quantum phase transition to a trivial phase, where the string order decreases continuously to zero from the colored density plot. To determine the exact transition point, we calculate the average particle number per site of the system. We illustrate our results for $J_R/J_L=1$ in Fig.~\ref{fig:number}, with the position of the abrupt change of particle number signaling this transition.

Before the transition point at around $g=2.2 $, the system belongs to the $\mathbb{Z}_3 \times \mathbb{Z}_3 $ symmetry-protected chiral Haldane phase, with the average particle numbers per site the same for each of quark spaces ($n_a=n_b=1/3$). Increasing anisotropy $g$ forces the system to polarize along $\vert a\rangle$, $\vert b\rangle$, or $\vert 0\rangle$ directions, and leads to the spontaneous breaking of one of the $\mathbb{Z}_3$ symmetries (the cyclic permutation of quark states) at the transition point. After this transition, the system polarizes at one of three quark spaces as shown in Fig.~\ref{fig:number}. The transition lines observed from this average particle number are shown as the solid lines in Fig.~\ref{fig:phasediagram}, and are consistent with the calculations of string order parameters.

\subsection{The variational ansatz}\label{ansatz}
\begin{figure}[h]
	\centering
	\includegraphics[width=0.8\linewidth]{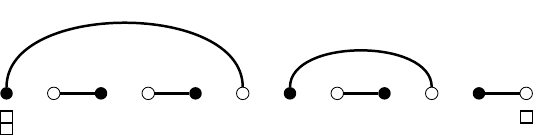}
	\caption{An illustration of higher-order singlet pairing in the ground state, where local quark state $\bm{3}$ can pair with anti-quark state $\overline{\bm{3}}$ at even longer distance. In our calculations, we only consider to lowest-order long-range pairing, that quark pairs with an anti-quark at its next-nearest neighbor (the long pairing line in the middle).}
	\label{fig:groundstate}
\end{figure}

To give a qualitative explanation of this spontaneous $\mathbb{Z}_3$ symmetry breaking, here we provide an analytical variational ansatz, and show the prediction of the transition points. We construct the ground state wavefunction from the fully connected valence bond solid (VBS) state, where each even or odd bond forms a SU(3) singlet. As proved by Affleck~\cite{affleck1985large}, this state is the exact ground state of the Heisenberg point in the large-$N$ limit. We then include long-range pairing states, as illustrated in Fig.~\ref{fig:groundstate}. For simplicity, in our calculation we only consider the lowest-order long-range pairing with single quark connecting to its next-nearest anti-quark. 

Considering the symmetry breaking from the left-chiral Haldane phase ($J_R/J_L\leq 1$), the variational ground state is written as
\begin{equation}\label{eq:psi}
	|\Psi\rangle =  c_0  |\phi\rangle ^{\otimes \frac{L}{2}} + c_1 \sum_{i=0}^{L/2-2}  |\phi\rangle ^{\otimes i} \otimes |\psi\rangle_{2i+1} \otimes  |\phi\rangle ^{\otimes \frac{L - 2i-4}{2}},
\end{equation}
where
\begin{equation}
	|\phi\rangle = \cos(\theta) |00\rangle + \frac{\sin(\theta)}{\sqrt{2}} \left( |aa\rangle + |bb\rangle \right)
\end{equation}
characterizes the local pairing of quark and anti-quark states, and the lowest-order long-range pairing is included in
\begin{align}
	&|\psi\rangle_{2i+1} = \cos(\theta)|0\rangle_{2i+1} \otimes |\phi\rangle \otimes |0\rangle_{2i+4}\nonumber\\
	&+\frac{\sin(\theta)}{\sqrt{2}} \left( |a\rangle_{2i+1} \otimes |\phi\rangle \otimes |a\rangle_{2i+4}+ |b\rangle_{2i+1} \otimes |\phi\rangle \otimes |b\rangle_{2i+4} \right)
\end{align}
which considers the pairing between the anti-quark state at site $2i+1$ and quark state at site $2i+4$. The construction of the right-chiral case is similar, and we provide its details in Appendix~\ref{rightpsi}.

\begin{figure}[h]
	\centering
	\includegraphics[width=0.8\linewidth]{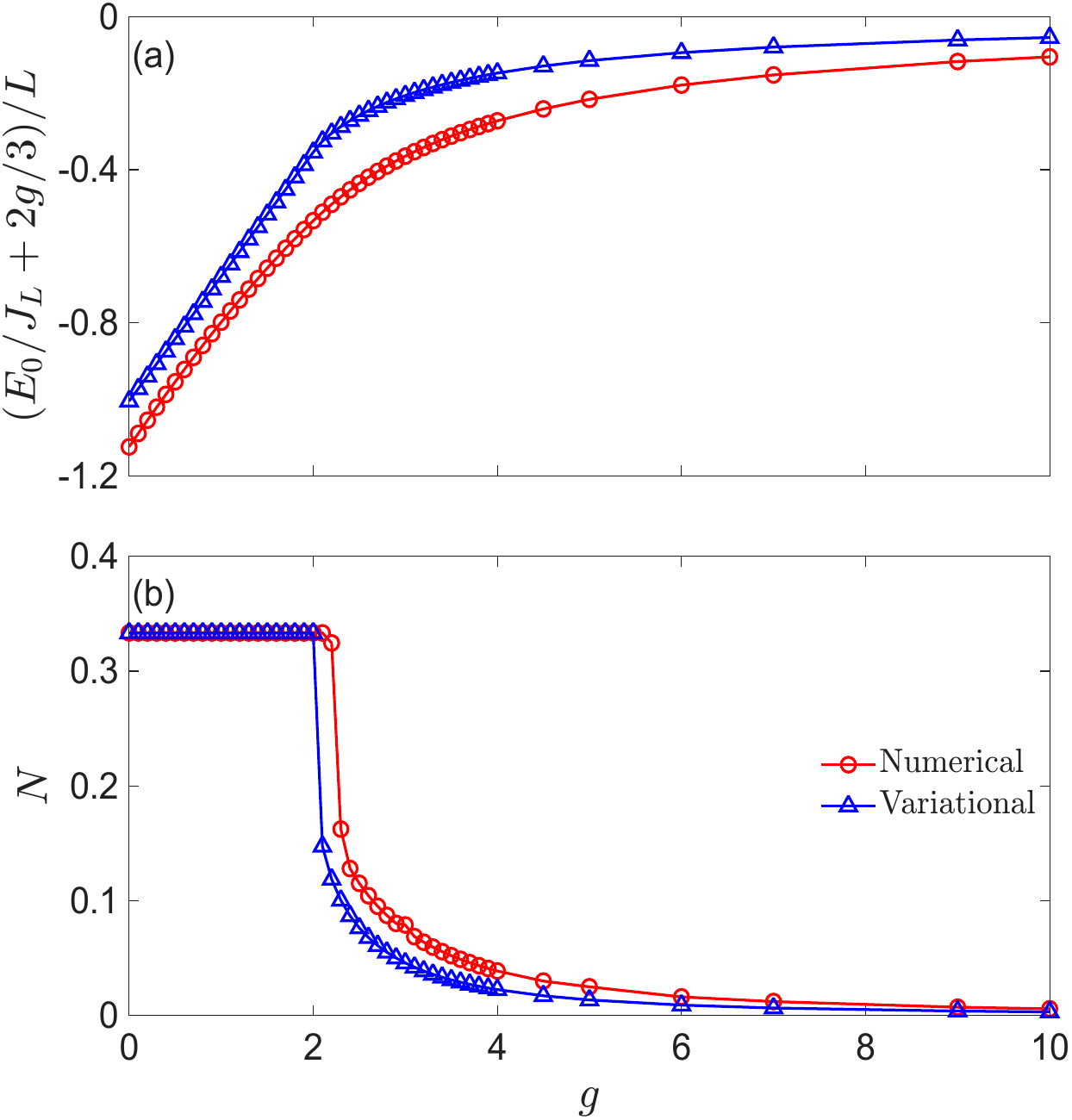}
	\caption{The ground-state energy in (a) and average particle number in (b) as a function of $g$ for $J_L=J_R$. The red line corresponds to our DMRG results (OBC, $ L = 180 $, $ m = 800 $), while the blue line is from our variational ansatz. Our variational calculations are consistent with numerical results, and predict a quantum phase transition at $ g \approx 2.1 $.}
	\label{fig:variation}
\end{figure}

From above ground state ansatz, we can calculate the corresponding variational energy. Note that different terms in our wavefunction Eq.~(\ref{eq:psi}) are not orthogonal, so the normalization condition should be carefully dealt with. We illustrate the results of our variational calculations for $J_R/J_L=1$ in Fig.~\ref{fig:variation}, and leave the somewhat lengthy expressions in Appendix~\ref{varenergy}.

From Fig.~\ref{fig:variation}, we can find our variational results give a very close ground state energy as the numerical calculations. Especially, it provides a very close prediction of the transition point from the average particle number. We show our variational prediction as the dashed lines in Fig.~\ref{fig:phasediagram}, and we can see it gives a consistent prediction as our numerical calculations.

\begin{figure}[b]
	\centering
	\includegraphics[width=0.8\linewidth]{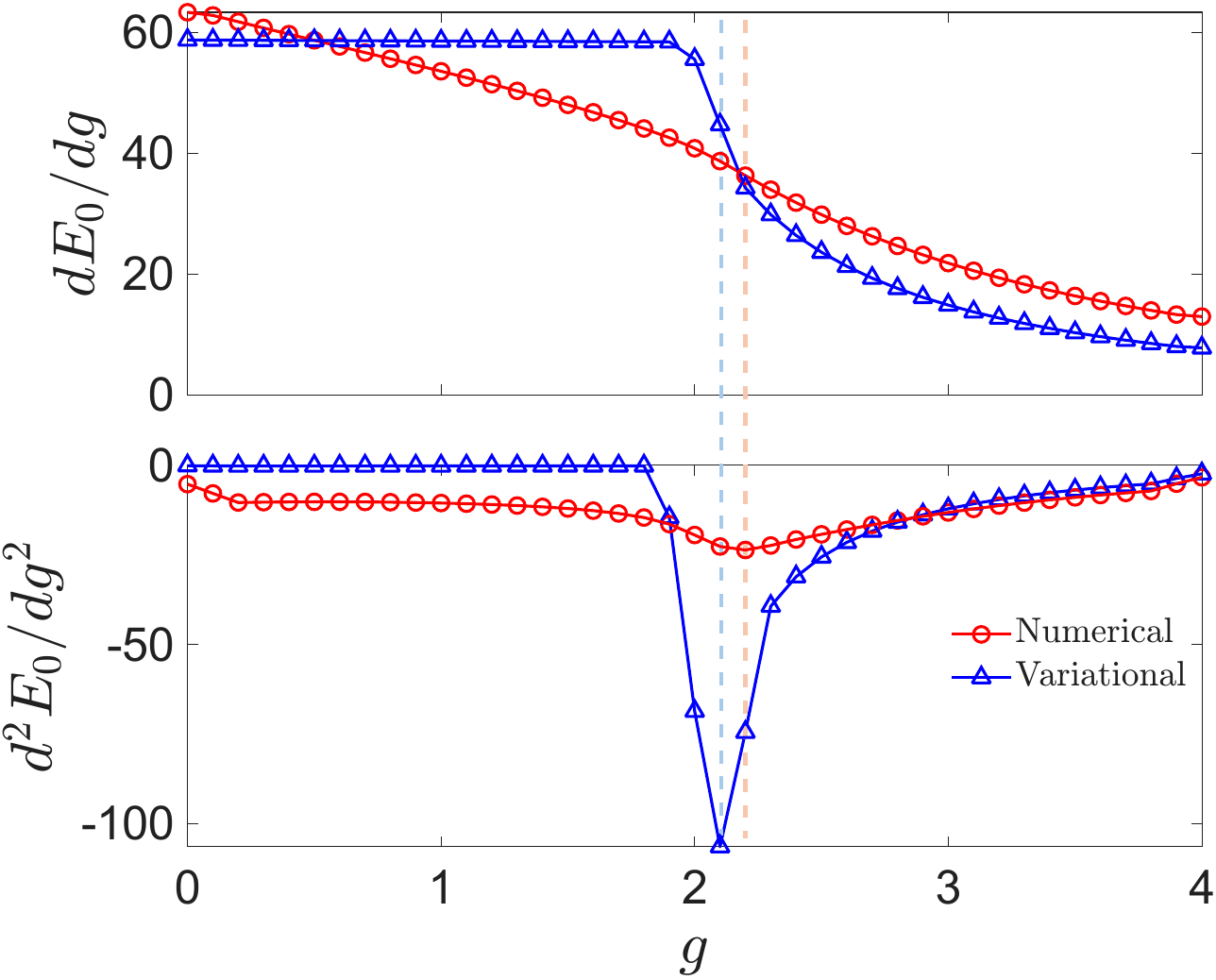}
	\caption{The first and second-order derivatives of the ground-state energy $E_0$ as a function of $g$ at $J_R/J_L=1$. The red curve shows the numerical result, while the blue curve gives our variational prediction. The transition points are identified by the minimum of the second derivatives (labeled as dashed lines), which gives around $g=2.2$ and $2.1$ for the numerical and variational calculations, respectively.}
	\label{fig:z3transition}
\end{figure}

\subsection{Nature of the phase transition}\label{z3transition}
We are now ready to identify the nature of this phase transition, especially at $J_R=J_L$. We calculate the first and second-order derivatives of the ground state energy at this point as shown in Fig.~\ref{fig:z3transition}. The transition points are labeled as the minimum of their second derivatives, which gives consistent predictions as our calculations of the average particle number in Fig.~\ref{fig:variation}. As we can see, in the first derivatives, both our numerical and variational calculations show very smooth changes at the transition points, especially for a larger $g$. Similar results are found away from the point $J_R/J_L=1$, and we leave the corresponding calculations to Appendix~\ref{engderivative}. Thus, we conclude that the $\mathbb{Z}_3$ symmetry-breaking phase transition here belongs to a continuous quantum phase transition.

\begin{figure}[t]
	\centering
	\includegraphics[width=.95\linewidth]{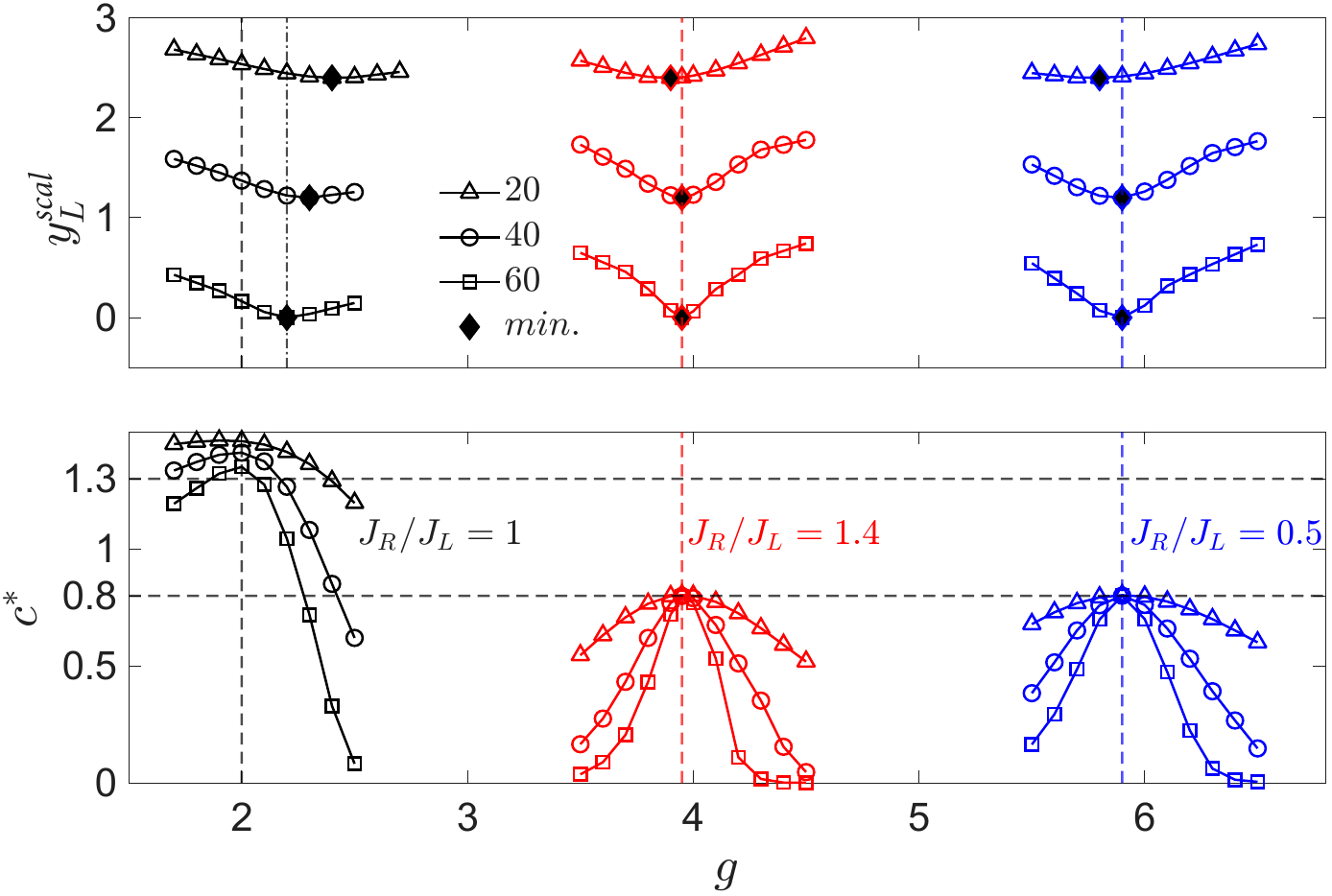}
	\caption{The central charge $c^{*}$ around the $\mathbb{Z}_3$ symmetry breaking transition at $J_R/J_L=1, 1.4, 0.5$ for system sizes $L=20, 40, 60$ (PBC). For comparison, we also show the rescaled second-order derivative of ground-state energy density $y_L^{scal} = \left(y_L - \min y_L\right)/|\min y_L|$ with $y_L(g) = (1/L)d^{2}E_0/dg^{2}$. For the transition starting from the left or right chiral ($J_R/J_L=0.5$ or $1.4$) state, the central charge quickly converges to a peak value at the critical point with $c^*=0.8$, which belongs to the 3-state Potts universality class. The locations of these critical points are consistent with the minimum of the second-order derivatives indicated by the vertical dashed lines. For the case $J_R/J_L=1$, the central charge seems to indicate a transition point at $g\approx 2$. However, we are unable to reach a convergent result, both for this central charge and the position of the minimum of the second-order derivative.}
	\label{fig:cstar}
\end{figure}

We further calculate the central charge of this phase transition. We use periodic boundary conditions, and compared to extracting the central charge from Eq. (\ref{eq:entropy}), it is found optimal to consider only the central bonds~\cite{nishimoto11, ejima14, ejima15}. Here, to avoid the effect of oscillations of the entanglement entropy due to dimerization, we calculate the central charge by
\begin{align}
	c^{*}(L) = \frac{3\bigl[S_L(L/2-2)-S_L(L/2)\bigr]}
	{\ln\!\bigl[\cos(2\pi/L)\bigr]} .
\end{align}
Figure \ref{fig:cstar} shows our results for the central charge $c^*$ around the $\mathbb{Z}_3$ symmetry breaking transition at three different values $J_R/J_L=1, 1.4, 0.5$, for system sizes $L=20, 40, 60$. For comparison, we also show the results of the rescaled second-order derivative of the ground-state energy density $y_L^{scal} = \left(y_L - \min y_L\right)/|\min y_L|$ with $y_L(g) = (1/L)d^{2}E_0/dg^{2}$. We find that for the transition starting from the left or right chiral ($J_R/J_L=0.5$ or $1.4$) phase, the peak of the central charge quickly converges to the value $c^*=0.8$ at the transition points (labeled by the vertical dashed lines). This signals a 3-state Potts universality class for this transition, and is also consistent with the critical behavior of the correlation function (see Appendix~\ref{engderivative}).  At $J_R/J_L=1$, this transition becomes more interesting. Besides the $\mathbb{Z}_3$ symmetry breaking, there is an additional $Z_2$ symmetry involved, due to the restoration of the translational invariance. The central charge seems to indicate a transition point at $g\approx 2$, which is consistent with our energy gap in Fig. \ref{fig:gap}. Unfortunately, we are unable to reach a convergent result, both for the central charge and the minimum of the second-order derivative. One might expect this transition to belong to the 3-state Potts + Ising universality class, which gives $c^*=0.8+0.5=1.3$  (see Appendix~\ref{multicritical}). We leave this as an open question, which might require further numerical techniques.

\section{The Experimental Proposal}\label{experiment}
\begin{figure}[h]
	\centering
	\includegraphics[width=0.9\linewidth]{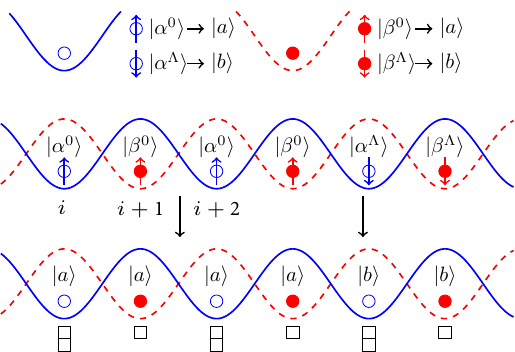}                            
	\caption{Illustration of our proposal for realizing the SU(3) AFH chain using two-species spin-1/2 bosons $\alpha$ and $\beta$. The spin state $|0\rangle$ and $|\Lambda\rangle$ correspond to the Holstein-Primakoff bosons $a$ and $b$, respectively. The atoms $\alpha$ and $\beta$ are confined in a species-dependent optical potential, so that the system can be stabilized under an interspecies attractive interaction.}
	\label{fig:exp}
\end{figure}

The most challenging part to realize our model is to implement a stable bosonic system with attractive nearest-neighbor interaction, while locally lying in the hardcore limit, as indicated by Eq.~(\ref{eq:fullham}). We propose that this can be realized in two-species spin-1/2 bosons that emerge from including $\Lambda$ and $V$ transitions in spinor $F=1$ Bose gases~\cite{krutitsky2004spin, larson2008coupled, de2010phase}. We illustrate our proposed system in Fig.~\ref{fig:exp}.

The two species $\alpha$ and $\beta$ are confined in a species-dependent optical lattice, with their spin states $\vert 0\rangle$ and $\vert \Lambda\rangle$ mapped to our local bosonic states $\vert a\rangle$ and $\vert b\rangle$, as shown in Fig.~\ref{fig:exp}. To realize the hardcore limit, we only need to consider the Hamiltonian of intraspecies spin-1/2 bosons as~\cite{krutitsky2004spin, de2010phase}:
\begin{align}
	H_\chi = & -t^\chi \sum_{\sigma,i} \left( \chi_{\sigma, i}^\dagger \chi_{\sigma,i+2} + \rm{H.c.}\right) + \frac{U_0^\chi}{2} \sum_{\sigma, i} \hat{n}_{\sigma, i}^\chi (\hat{n}_{\sigma, i}^\chi - 1)\nonumber \\
	& + \frac{U_2^\chi}{2} \cos(\delta \phi^\chi) \sum_i \left( \chi_{0,i}^\dagger \chi_{0,i}^\dagger \chi_{\Lambda, i} \chi_{\Lambda, i} + \rm{H.c.} \right) \nonumber\\
	& + (U_0^\chi + U_2^\chi) \sum_i \hat{n}_{0,i}^\chi \hat{n}_{\Lambda, i}^\chi
		- \mu^\chi \sum_{\sigma, i} \hat{n}_{\sigma, i}^\chi,
	\label{eq:hamchi}
\end{align}
where $\chi=\alpha, \beta$ label the two species with $\sigma=0, \Lambda$ corresponding to their spin components. The relative phase $\delta \phi^\chi=2(\phi^\chi_0-\phi^\chi_\Lambda)$ measures the difference between the global phases of $\vert 0\rangle$ and $\vert \Lambda\rangle$ states. For the intraspecies interaction here, this term is trivial, but it will become clear later that this relative phase plays a crucial role in our proposal.

The spin-independent interaction $U_0^\chi$ and spin-dependent one $U_2^\chi$ are proportional to the scattering lengths as~\cite{kawaguchi2012spinor}:
\begin{equation}
	U^\chi_0 \propto \frac{a^\chi_0 + 2 a^\chi_2}{3}, 
	\quad 
	U^\chi_2 \propto \frac{a^\chi_2 - a^\chi_0}{3},
\end{equation}
where $a^\chi_F$ represents the $s$-wave scattering length of total spin $F$ channel. Thus, to realize the hardcore limit of Hamiltonian Eq.~(\ref{eq:hamchi}), we can tune the Feshbach resonance to reach a large positive $a^\chi_2$,  so that both interaction terms with strengths $U_0^\chi/2$ and $U_0^\chi+U_2^\chi$ dominate,  and two $\alpha$ or $\beta$ particles at the same site are strongly repulsive~\cite{hamley2009photoassociation,papoular2010microwave}.

The most important part would be the attractive interaction between $\alpha$ and $\beta$ particles at nearest neighbor sites. By generalizing Hamiltonian Eq.~(\ref{eq:hamchi}) to nearest sites, we find
\begin{align}
	H_{\alpha \beta} = &
		\frac{U_0^{\alpha\beta}}{2} \sum_{\sigma, i} n_{\sigma, i}^{\chi} n_{\sigma, i+1}^{\chi'} \nonumber\\ 
	& + \frac{U_2^{\alpha\beta}}{2} \cos(\delta \phi^{\alpha\beta}) \sum_{i} \left(
		\chi_{0, i}^\dagger \chi_{0, i+1}^{\prime \dagger} \chi_{\Lambda, i} \chi_{\Lambda, i+1}^{\prime}+\rm{H.c.}\right)\nonumber\\
	& + U_n^{\alpha\beta}  \sum_{i} \left(
		n_{0, i}^{\chi} n_{\Lambda, i+1}^{\chi'} + n_{\Lambda, i}^{\chi} n_{0, i+1}^{\chi'} \right)\nonumber\\
	& + U_c^{\alpha\beta}  \sum_{i} \left(
		\chi_{0, i}^\dagger \chi_{\Lambda, i+1}^{\prime \dagger} \chi_{\Lambda, i} \chi_{0, i+1}^{\prime}+\rm{H.c.} \right)
\end{align}
with $ \chi = \alpha $ or $ \beta $ and $ \chi \neq \chi' $. The interaction strengths
\begin{eqnarray}
	U_0^{\alpha\beta} \propto \frac{a^{\alpha\beta}_0 + 2 a^{\alpha\beta}_2}{3}, 
	\quad 
	U_2^{\alpha\beta} \propto \frac{a^{\alpha\beta}_2 - a^{\alpha\beta}_0}{3},\nonumber\\
	U_n^{\alpha\beta} \propto \frac{a^{\alpha\beta}_2 + a^{\alpha\beta}_1}{4}, 
	\quad 
	U_c^{\alpha\beta} \propto \frac{a^{\alpha\beta}_2 - a^{\alpha\beta}_1}{4},
\end{eqnarray}
where $a^{\alpha\beta}_F$ corresponds to the interspecies $s$-wave scattering lengths at total spin $F$ channel.

Compared with the Hamiltonian Eq.~(\ref{eq:fullham}), to realize the Heisenberg point one needs to tune $U^{\alpha\beta}_c=0$ and $U^{\alpha\beta}_0=4U^{\alpha\beta}_n<0$ , which means $a^{\alpha\beta}_0 = 4 a^{\alpha\beta}_1$ and $a^{\alpha\beta}_1=a^{\alpha\beta}_2 < 0$.  However, this leads to a positive interaction strength $U^{\alpha\beta}_2/2$, in contrast to our Hamiltonian Eq.~(\ref{eq:fullham}). Thanks to the appearance of the relative phase, studies show that this system will choose a finite phase difference $\delta\phi^{\alpha\beta}=2(\phi^\alpha_0-\phi^\beta_\Lambda)=\pm\pi$ for $U^{\alpha\beta}_2>0$~\cite{krutitsky2004spin}. This makes sure that our color-converting term $U^{\alpha\beta}_2\cos(\delta\phi^{\alpha\beta})/2$ is always negative, which is crucial in stabilizing the singlet bonds that are needed in the topological phases.

We then mention that the pair creation and annihilation terms that appear in the Hamiltonian Eq.~(\ref{eq:fullham}) can be implemented via a two-photon Raman photoassociation process~\cite{rom2004state, gupta2010effect, dutta2017two}, which couples the interspecies $\Lambda-\Lambda$ and $0-0$ pairs into higher level molecular states with total spin $F=0$. Additionally, the chiral topological phases can be revealed by including an additional lattice potential that shifts the position of these two species-dependent lattices.

At last, we discuss the stability of our proposed system. As we have mentioned, the main obstacle here is how to create a stable bosonic system with attractive interaction. In our proposed two-lattice structure, as shown in Fig.~\ref{fig:exp}, the on-site interaction is strongly repulsive. The only attractive interaction comes from interspecies, which happens due to the overlap of the corresponding Wannier functions that appear mostly in the middle of two traps. This makes sure that the attractive interaction is greatly reduced and thus stabilizes the bosonic system.

\section{Summary}\label{summary}
In this work, we provide a Holstein-Primakoff bosons realization of the SU(3) alternating conjugate representation AFH chains. By mapping the quark spaces into hardcore bosonic states, we hope to realize such a model in the highly tunable cold atomic system. To explore the chiral topological phases and their transitions, we focus on two parameters, the interaction ratio between even and odd bonds $J_R/J_L$, and the anisotropy $g$ along the $T^3$ and $T^8$ directions. 

We identify the left and right-chiral topological phases with a chiral-reversed topological transition at $J_R=J_L$. We confirm that this quantum phase transition belongs to the first order, which supports a finite energy gap and nontrivial topological string orders. Most importantly, we show that around the Heisenberg point, there is a rather small region where not only the ground state, but also the first excited state shows topological behavior. This finding provides us with an alternative mechanism to support SPT states at higher energy levels.

We then find a spontaneous $\mathbb{Z}_3$ symmetry breaking by increasing the anisotropy $g$. We provide a variational wavefunction ansatz, and through the average particle number, we give a reasonable prediction of this topological to trivial quantum phase transition. We also provide an experimental system to realize this model by loading two-species hardcore spin-$1/2$ bosons into a species-dependent optical lattice.

Finally, we mention the current experiment. Both the site-resolved occupation numbers and the nonlocal string order parameters can be measured using quantum gas microscopes~\cite{bakr2009quantum, bakr2010probing, sherson2010single, endres2011observation, sompet2022realizing}. The main experimental challenge might lie in tuning the multiple optical or microwave Feshbach resonances. While identifying a suitable bosonic species remains an open task, there are already several well-studied heterospecies spin-1 bosonic systems (e.g. $^{23}{\rm Na}-^{87}{\rm Rb}$~\cite{li2015coherent,li2020manipulation} and $^7{\rm Li}-^{87}{\rm Rb}$~\cite{fang2020collisional}).

\section*{Acknowledgements}
This research is supported by the Fundamental Research Funds for the Central Universities (No. FRF-BD-25-033).

\appendix
\section{Reveal the edge states}\label{edge}
\begin{figure}[ht]
	\centering
	\includegraphics[width=1.0\linewidth]{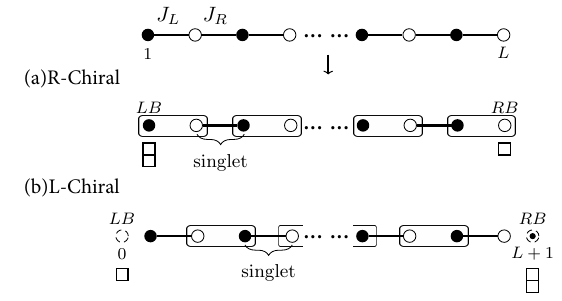}
	\caption{Schematic illustration of the edge states in the (a) right and (b) left chiral phases. We choose different unit cells (shown as the black boxes) in these two phases, and label their left and right edge states (indicated as LB or RB). For the right chiral state, the edge states can be clearly revealed, while for the left one, we need special techniques to reveal its edge excitations (see the LB and RB).}
	\label{fig:edgeproject}
\end{figure}

We characterize the topological edge states by the local operator ${\cal O}^{R/L}_i$ that are defined in different two-site unit cells in Eq.~(\ref{eq:unitop}) (see Fig.~\ref{fig:edgeproject} for a schematic view). This projects our system to a virtual one with local adjoint representation. As illustrated in Fig.~\ref{fig:edgeproject}, for the right chiral phase, the edge states in the virtual system appear naturally in the actual physical system (see the LB and RB in Fig.~\ref{fig:edgeproject}a). However, for the left chiral one, its virtual edge excitations are screened out by residual edges, and thus we need special techniques to reveal its virtual topological excitations.

\begin{figure}[h]
	\centering
	\includegraphics[width=0.8\linewidth]{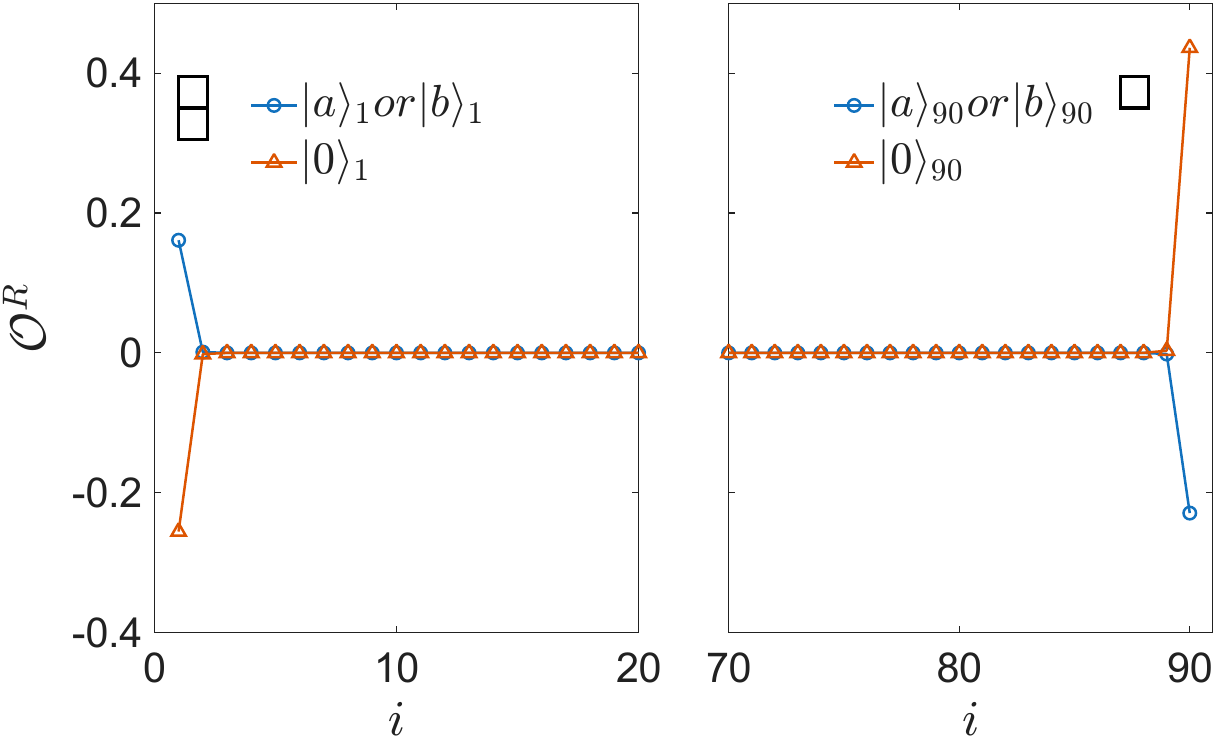}
	\caption{Different edge states that appear in the nine-fold right chiral states, while for each state we observe a single anti-quark state on the left boundary, and a quark state on the right boundary. The parameters are chosen as $g=1$ and $J_R/J_L=32$ ($L=180$, $m=800$), and we show the unit cells.}
	\label{fig:fulledges}
\end{figure}

\begin{figure}[h]
	\centering
	\includegraphics[width=0.8\linewidth]{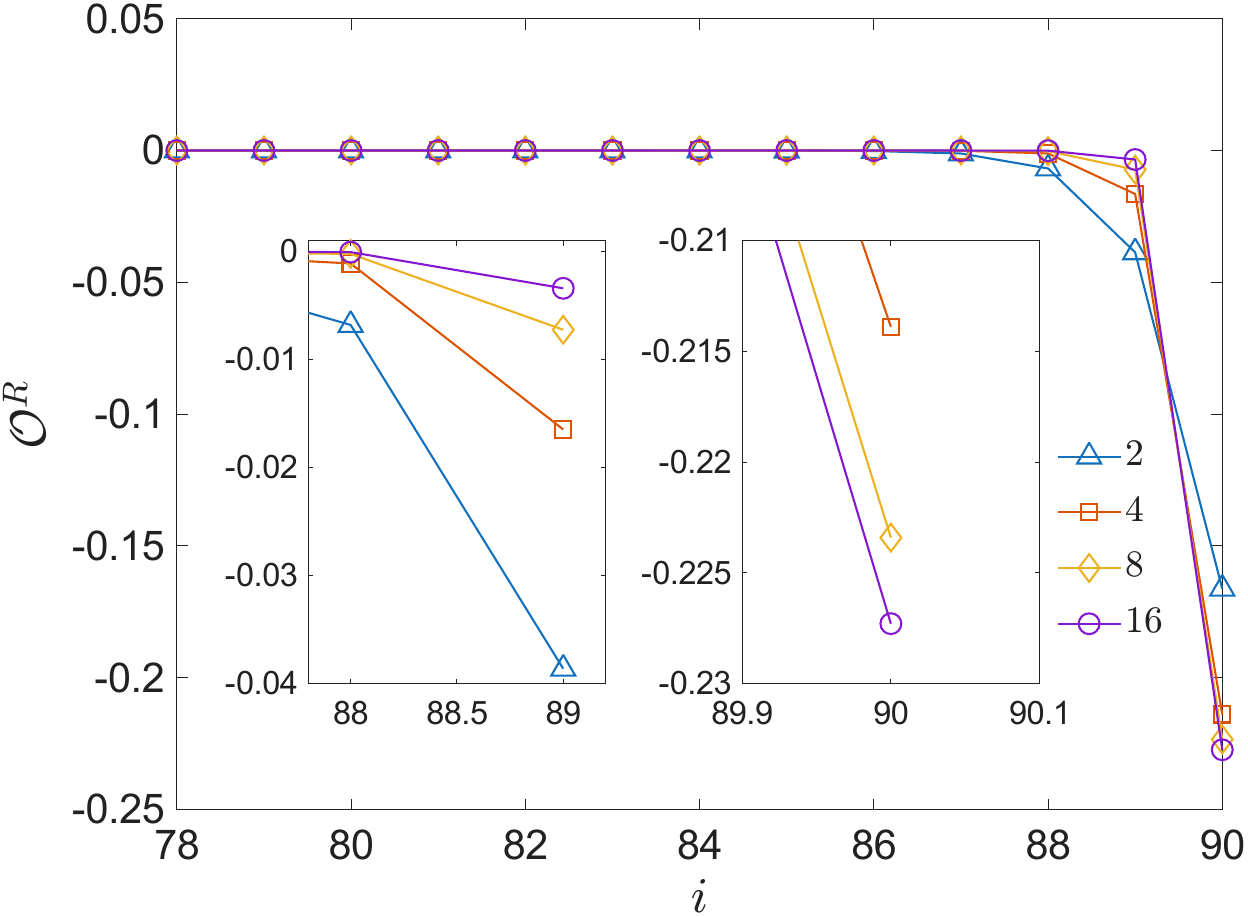}
	\caption{The right chiral edge states for various $J_R/J_L=2,\,4,\,8,\,16$ ($g=1$, $L=180$, $m=800$). We only show the right boundaries that support edge states $\lvert b\rangle$. The insets give enlarged views of the regions around unit cells 88--89 and 90.}
	\label{fig:RCedge}
\end{figure}

For the right chiral states, we show their topological edge states on both boundaries in Fig.~\ref{fig:fulledges}. We note that the open boundary system supports nine-fold degeneracy in its ground states, we show all of their edge states in the virtual unit cells, and label its quark or anti-quark states at $g=1$ and $J_R/J_L=32$. We also show the evolution of the right edge states as a function of coupling $J_R/J_L=2, 4, 8, 16$ in Fig.~\ref{fig:RCedge}.

\begin{figure}[h]
	\centering
	\includegraphics[width=0.8\linewidth]{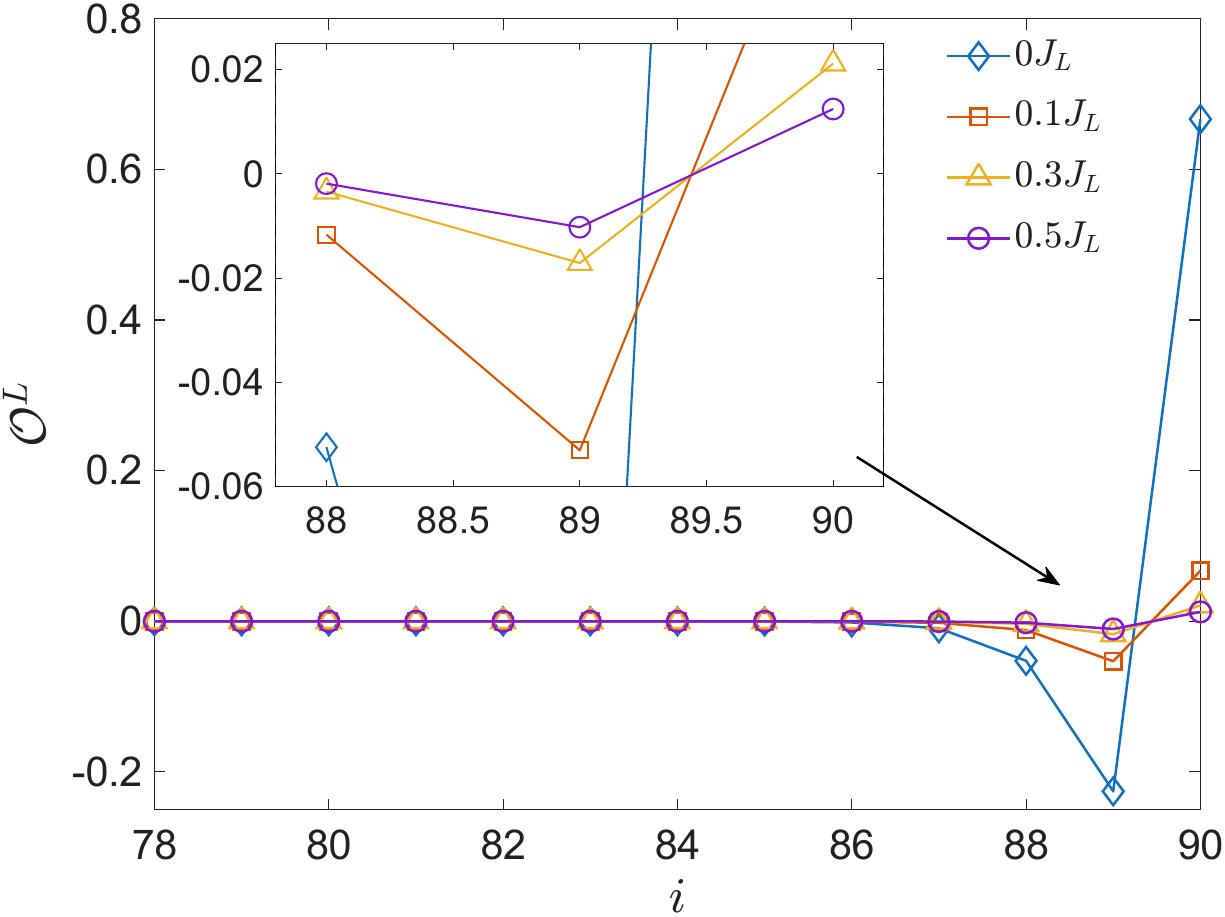}
	\caption{Reveal the edge states in the left chiral phase by reducing its boundary coupling $J_{179,180}/J_L=0,\,0.1,\,0.3,\,0.5$. The parameters are chosen as $g=1$, $J_R/J_L=0.5$, $L=180$, and $m=800$. The inset shows an enlarged view of sites 88--90 near the right edge.}
	\label{fig:LCedgecut}
\end{figure}

\begin{figure}[h]
	\centering
	\includegraphics[width=0.8\linewidth]{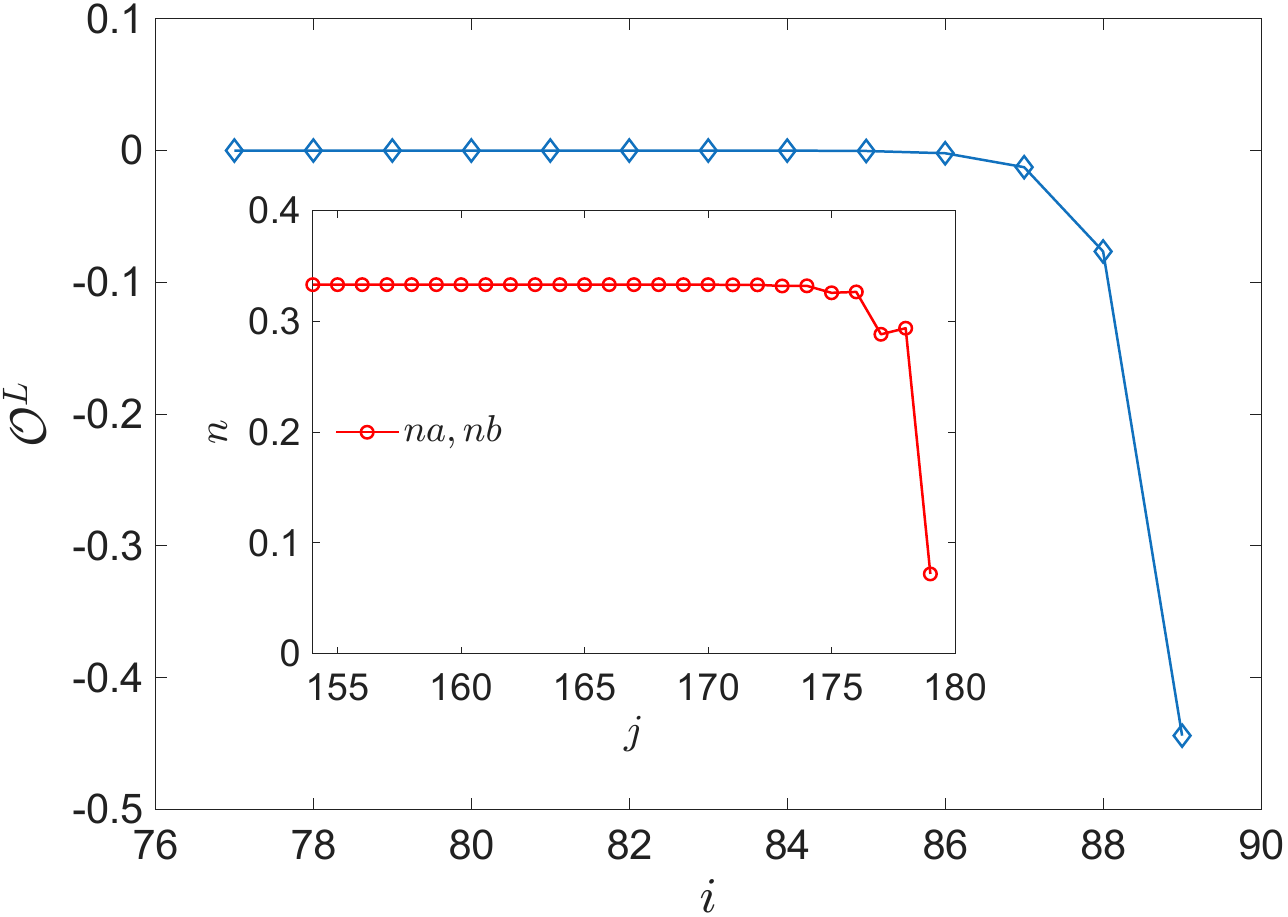}
	\caption{Reveal the edge states of the left chiral phase by pinning the boundary. We introduce a large local chemical potential on the site $180$ to pin its edge at state $\lvert 0\rangle$, and site $179$ serves as the right boundary of the projected system. The parameters are chosen as $g=1$, $J_R/J_L=0.5$, $L=180$, and $m=800$. We illustrate the corresponding edge states in both the unit cells and their physical lattice sites (see the inset).}
	\label{fig:LCedgepin}
\end{figure}

For the left chiral states, we provide two strategies to reveal their virtual topological excitations. The first way is more straightforward, by just cutting these residual edge sites. Our results are shown in Fig.~\ref{fig:LCedgecut}, and we reduce the boundary coupling between the sites $179$ and $180$ for $J_{179,180}=0,0.1,0.3,0.5$. This is just equivalent to Fig.~\ref{fig:edgeproject}a, and we shave the boundary to access its virtual edge. As we can see, the virtual topological excitations at unit cell $i=89$ become apparent. The other way is more experimentally accessible, that is, by pinning the actual physical boundary. We show the corresponding results in Fig.~\ref{fig:LCedgepin}. We include a large chemical potential at the right boundary, so that the physical right boundary site is pinned at state $\lvert 0\rangle$. In this way, the topological edge excitations at the virtual system become experimentally accessible. We show both the virtual and physical lattice sites, and the topological excitations appear at both sites.

\section{Ground-state ansatz for right-chiral symmetry breaking}\label{rightpsi}
For the spontaneous symmetry breaking of right-chiral Haldane phases, our construction of the ground-state wavefunction is similar to our left one Eq.~(\ref{eq:psi}), and written as
\begin{align}
|\Psi_{r}\rangle = &  c_a | e \rangle \otimes |\phi\rangle ^{\otimes \frac{L}{2}-1} \otimes | e \rangle\nonumber \\
&+ c_b \sum_{i=0}^{L/2-3}  | e \rangle \otimes|\phi\rangle ^{\otimes i} \otimes |\psi\rangle_{2i+2} \otimes  |\phi\rangle ^{\otimes \frac{L - 2i-6}{2}} \otimes | e \rangle,
\end{align}
where
\begin{align}
	 | e \rangle=\cos(\theta) |0\rangle + \frac{\sin(\theta)}{\sqrt{2}} \left( |a\rangle + |b\rangle \right)
\end{align}
characterizes the edge modes in open boundaries, and
\begin{equation}
	\begin{aligned}
		&|\psi\rangle_{2i+2} = \cos(\theta)|0\rangle_{2i+2} 
		\otimes |\phi\rangle \otimes |0\rangle_{2i+5}\\ &+\frac{\sin(\theta)}{\sqrt{2}} \left( |a\rangle_{2i+2} 
		\otimes |\phi\rangle \otimes |a\rangle_{2i+5}+ |b\rangle_{2i+2} \otimes |\phi\rangle \otimes |b\rangle_{2i+5} \right)
	\end{aligned}
\end{equation}
includes a single lowest-order long-range pairing. Note that in our calculations, we neglect the influence of boundaries and consider the coefficient $c_b$ to be site-independent. This approximation is reasonable, as long as we consider the energy of a large enough chain length.

\section{Expressions of the variational energy}\label{varenergy}
Here, we provide an explicit expression of the variational energy that characterizes left-chiral symmetry breaking. By plugging our ground-state wavefunction Eq.~(\ref{eq:psi}) to the Hamiltonian Eq.~(\ref{eq:fullham}), we get 
\begin{equation}
\langle \Psi | H | \Psi \rangle =\langle \Psi | H_\text{diag} | \Psi \rangle + \langle \Psi | H_\text{off} | \Psi \rangle,
\end{equation}
where
\begin{widetext}
	\begin{align}
		\langle \Psi | H_\text{diag} | \Psi \rangle =\;& g J_R A(t) \left( \frac{N}{2} - 1 \right) a^2  + g A(t) \left\{\left( \frac{N}{2} - 1 \right)J_L A(t)+  \left( \frac{N}{2} - 1 \right) \left( \frac{N}{2} - 2 \right)J_R\left( A(t) + C(t) \right) \right.\nonumber\\ 
		&\left. +  \left( \frac{N}{2} - 2 \right) \left( \frac{N}{2} - 3 \right) \left[\left( \frac{N}{2} - 3 \right) J_R C(t)^2  + 2 J_R D(t) \right] \right\}  b^2 + 2 g J_R A(t) C(t) \left( \frac{N}{2} - 2 \right) \left( \frac{N}{2} - 1 \right) a b, \\
		\langle \Psi | H_\text{off} | \Psi \rangle =\;& -2 J_L B(t) \frac{N}{2} a^2 \nonumber \\
		& -2 B(t) \left\{ \left( \frac{N}{2} - 1 \right) \left[ \left( \frac{N}{2} - 2 \right)J_L + J_R \right] + \frac{N}{2} J_L \left[ C(t)^2 \left( \frac{N}{2} - 2 \right)  \left( \frac{N}{2} - 3 \right) + 2 D(t) \left( \frac{N}{2} - 2 \right) \right] \right\} b^2 \nonumber \\
		& -\left\{ \left(\frac{N}{2} -1 \right) \left[ 4 B(t)  \left( \frac{N}{2} -2 \right) J_L C(t) + 6 \bigg( \sqrt{2} \sin t \cos^3 t + \frac{1}{\sqrt{2}} \sin^3 t \cos t 
		+ \frac{1}{2} \sin^4 t \bigg) J_R \right] \right\}ab,
	\end{align}
\end{widetext}
with the coefficients defined as
\begin{align}
	A(t) &= 2 \sin^2 t \cos^2 t + \frac{1}{2} \sin^4 t, \nonumber \\
	B(t) &= \sqrt{2} \sin t \cos t + \frac{1}{2} \sin^2 t, \nonumber \\
	C(t) &= \cos^4 t + \frac{1}{2} \sin^4 t, \nonumber \\
	D(t) &= \cos^6 t + \frac{1}{4} \sin^6 t.
\end{align}

\begin{figure}[b]
	\centering
	\includegraphics[width=0.8\linewidth]{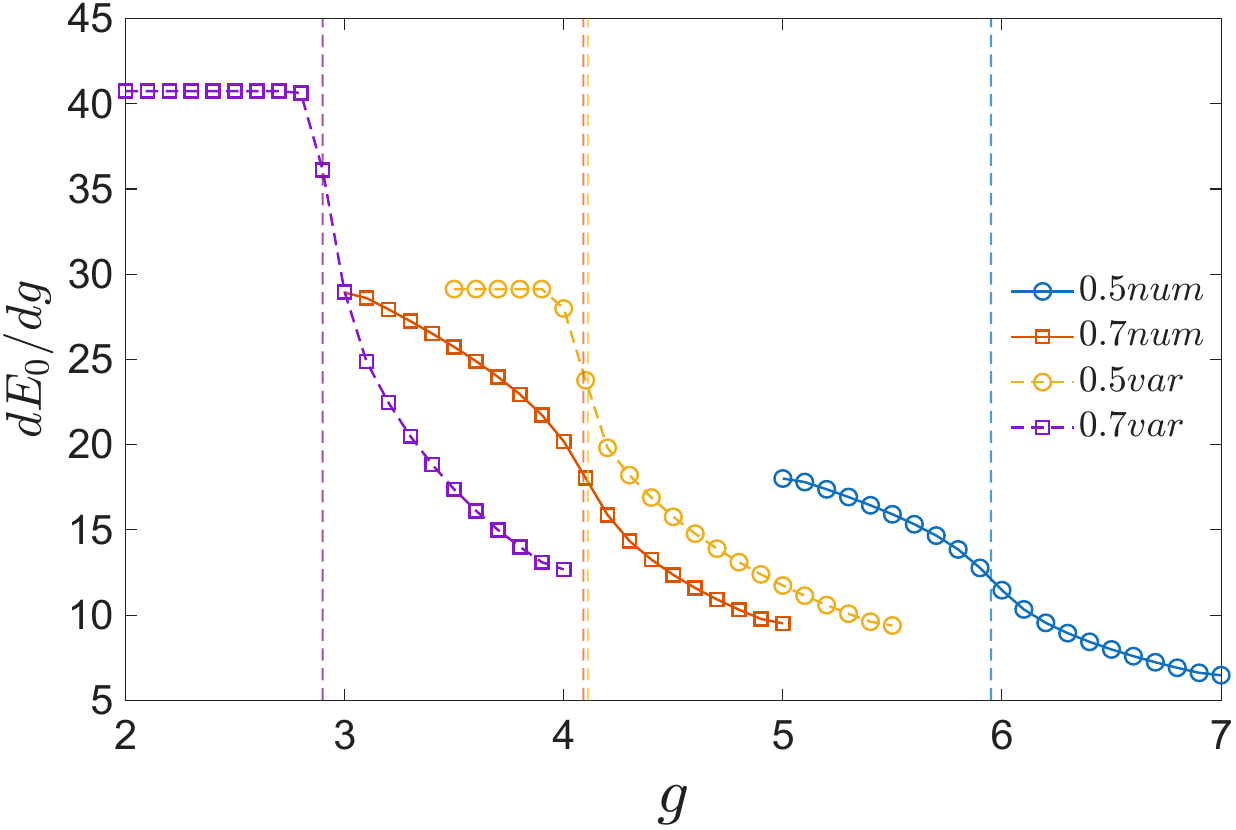}
	\caption{The first-order derivative of the ground-state energy from the left chiral side at $J_R/J_L = 0.5, 0.7$ (OBC, $L = 180 $, $ m = 800 $). The solid curves show the numerical results, while the dashed curves give our variational results. The vertical dashed lines correspond to the transition points identified in Fig.~\ref{fig:lc2nd}.}
	\label{fig:lc1st}
\end{figure}

As mentioned in Sec.~\ref{ansatz}, the states that constitute our variational ansatz are not orthogonal, so we need to renormalize our energy by the following condition
\begin{align}
	\langle \Psi | \Psi \rangle =\;& c_0^2 
	+ \left[ \left( \frac{N}{2} - 1 \right) 
	+ C(t)^2 \left( \frac{N}{2} - 2 \right) \left( \frac{N}{2} - 3 \right) \right. \nonumber \\
	& \left. + 2 D(t) \left( \frac{N}{2} - 2 \right) \right] c_1^2
	+ 2 C(t) \left( \frac{N}{2} - 1 \right) c_0 c_1.
\end{align}

\begin{figure}[hb]
	\centering
	\includegraphics[width=0.8\linewidth]{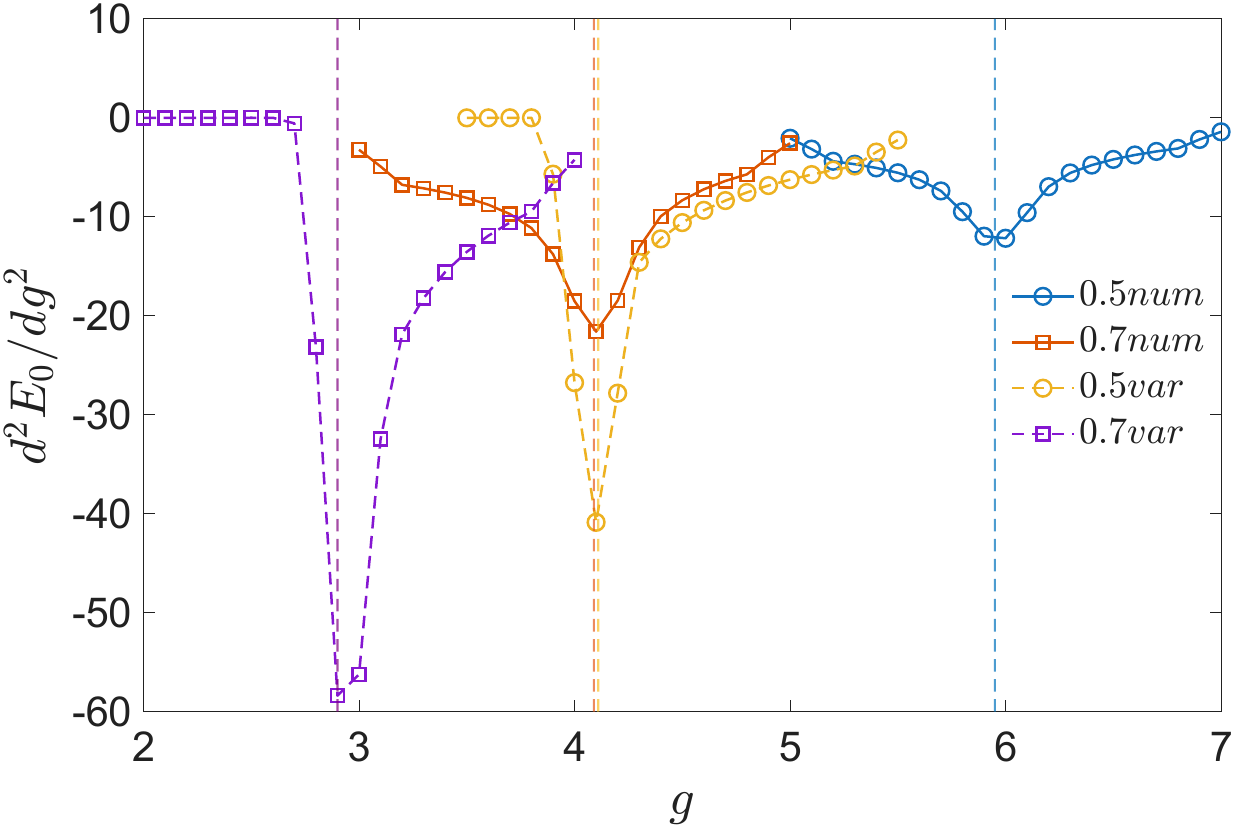}
	\caption{The second-order derivative of the ground-state energy from the left chiral side at $J_R/J_L = 0.5, 0.7$ (OBC, $L = 180 $, $ m = 800 $). The solid curves show the numerical results, while the dashed curves give our variational results. We label the transition points by the minimum of the corresponding second derivatives, as shown by the vertical dashed lines.}
	\label{fig:lc2nd}
\end{figure}

\section{Identify the phase transition from the chiral SPT to $\mathbb{Z}_3$ symmetry broken phase}\label{engderivative}
\begin{figure}[t]
	\centering
	\includegraphics[width=0.8\linewidth]{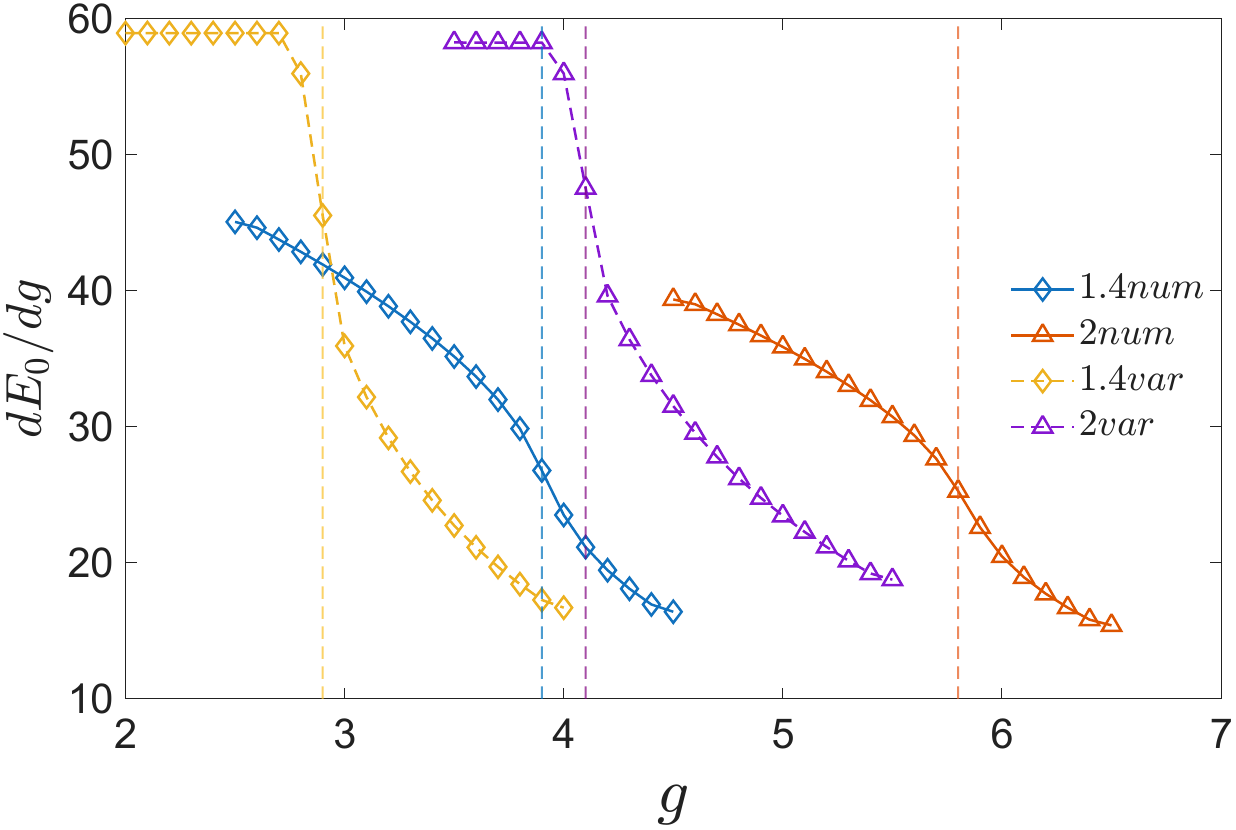}
	\caption{The first-order derivative of the ground-state energy from the right chiral side at $J_R/J_L = 1.4, 2.0$ (OBC, $L = 180 $, $ m = 800 $). The solid curves show the numerical results, while the dashed curves give our variational results. The vertical dashed lines correspond to the transition points identified in Fig.~\ref{fig:rc2nd}.}
	\label{fig:rc1st}
\end{figure}

\begin{figure}[ht]
	\centering
	\includegraphics[width=0.8\linewidth]{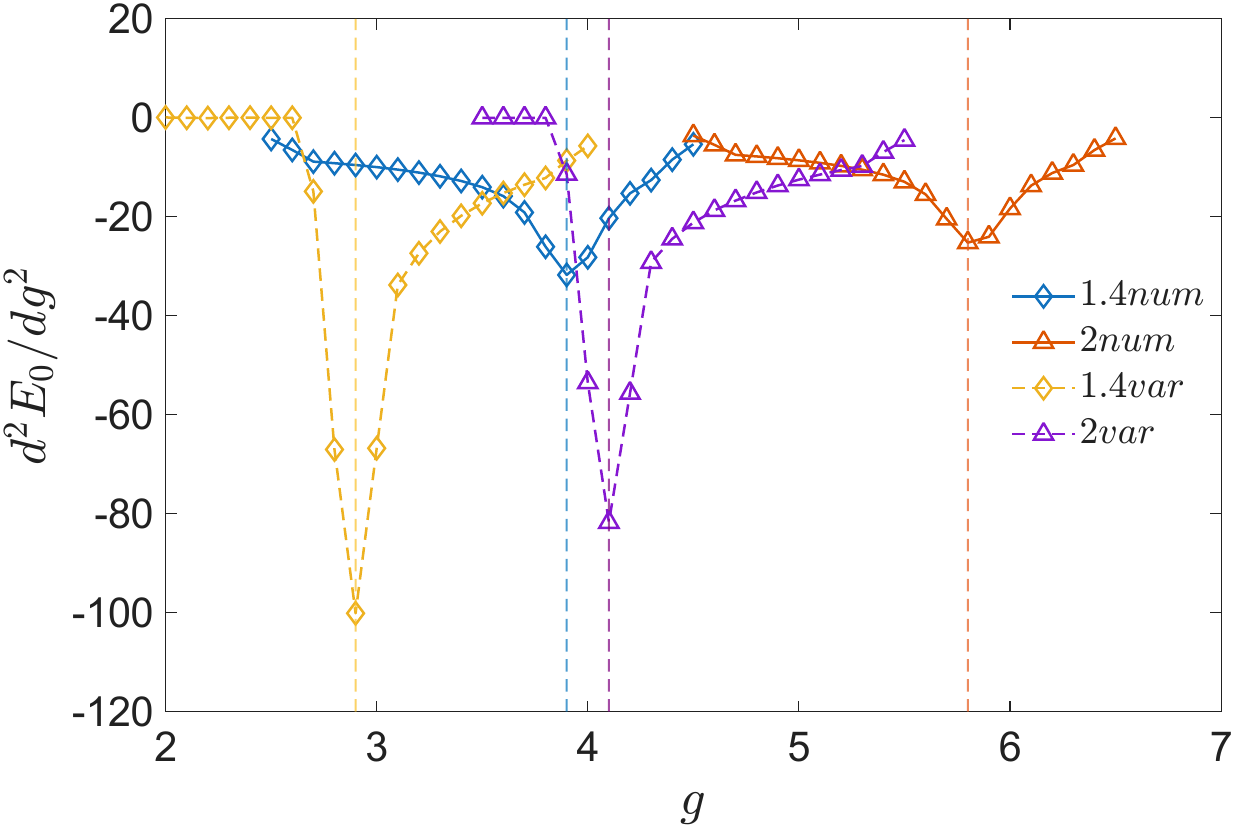}
	\caption{The second-order derivative of the ground-state energy from the right chiral side at $J_R/J_L = 1.4, 2.0$ (OBC, $L = 180 $, $ m = 800 $). The solid curves show the numerical results, while the dashed curves give our variational results. We label the transition points by the minimum of the corresponding second derivatives, as shown by the vertical dashed lines.}
	\label{fig:rc2nd}
\end{figure}

We show our calculations of the first and second-order derivatives of the ground-state energy from the left (Fig.~\ref{fig:lc1st}--\ref{fig:lc2nd}) and right (Fig.~\ref{fig:rc1st}--\ref{fig:rc2nd}) chiral sides. Consistent with our results in Sec.~\ref{z3transition}, we find that both our numerical and variational results support a continuous quantum phase transition, with the first derivatives changing smoothly at the transition points. We also observe larger deviations of the transition points between our variational prediction and the numerical one, which might be due to the higher-order long-range pairings that we have ignored in our variational ansatz, especially for a larger $g$.

\begin{figure}[t]
	\centering
	\includegraphics[width=0.8\linewidth]{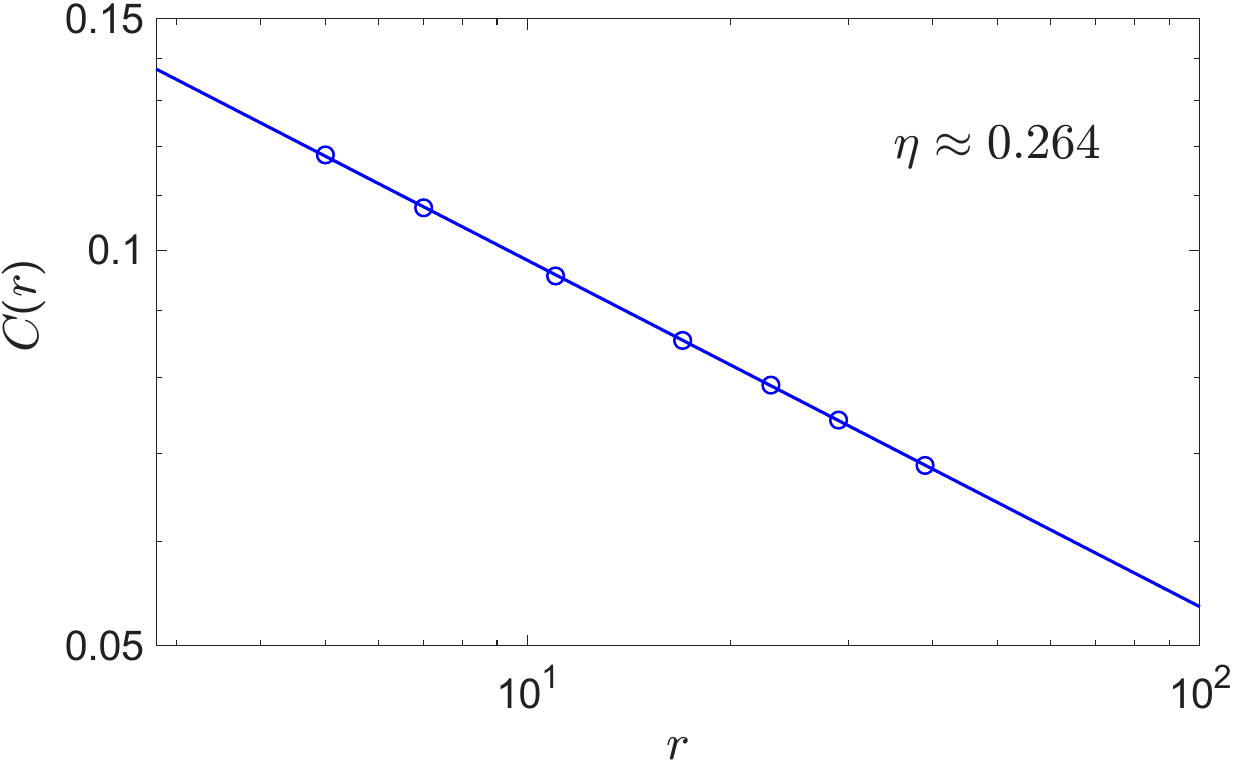}
	\caption{Critical behavior of the correlation function $C(r)$ at the $\mathbb{Z}_3$ symmetry breaking transition point $g\approx 5.9$ for $J_R/J_L=0.5$. Our fitting gives an anomalous dimension $\eta\approx 0.264$, consistent with the exact value $\eta=4/15$ predicted by the 3-state Potts universality class.}
	\label{fig:eta}
\end{figure}

We further calculate the critical behavior of the correlation function $C(r)$ at this $\mathbb{Z}_3$ symmetry breaking transition. Our results are illustrated in Fig. \ref{fig:eta}, where we choose $J_R/J_L=0.5$ with the transition point $g\approx 5.9$. The correlation function is well described by the scaling law, with the fitting giving an anomalous dimension $\eta\approx 0.264$, signaling a 3-state Potts universality class with the exact value $\eta=4/15$. This result is consistent with our central charge analysis in the main text.

\section{The central charge at $J_R/J_L=1$  $\mathbb{Z}_3$ transition point}\label{multicritical}
\begin{figure}[t]
	\centering
	\includegraphics[width=0.8\linewidth]{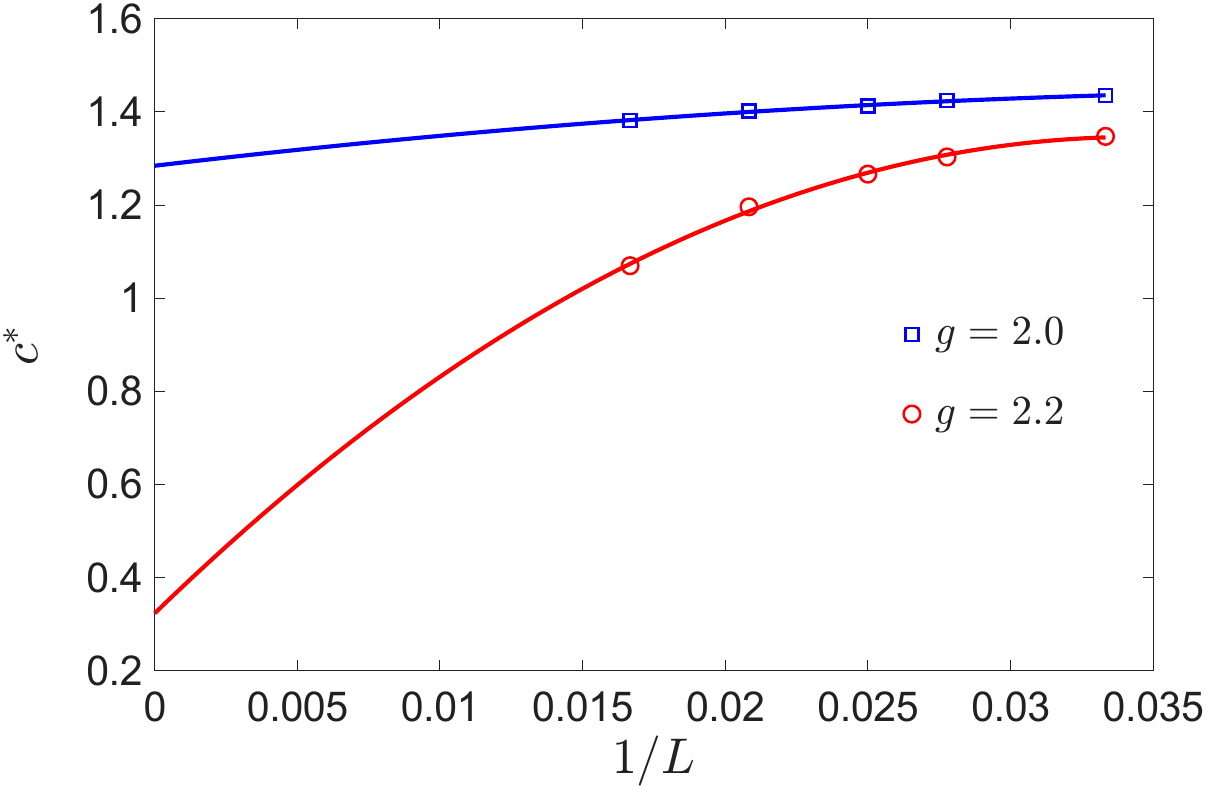}
	\caption{The central charge $c^{*}$ as a function of the system size $1/L$ at $g = 2$ for $J_R/J_L = 1$. We do a second-order polynomial fitting as the blue solid line, which gives around $c^*\approx 1.3$ for $g=2$. For comparison, we also show the results for $g=2.2$, with the fitting giving $c^{*} \approx 0.3$, well below the value of 3-state Potts or Ising universality class.}
	\label{fig:cstarfit}
\end{figure}

We do a second-order polynomial fitting of the central charge $c^*$ at the $J_R/J_L=1$  $\mathbb{Z}_3$ transition point with limited system sizes. As illustrated in Fig. \ref{fig:cstarfit}, our fitting gives a rather close value to $c^*\approx 0.8+0.5=1.3$ at $g=2$, suggesting this transition point might belong to the 3-state Potts + Ising universality class. For comparison, the fitting at $g=2.2$ gives $c^*\approx 0.3$, which is well below the values predicted by the 3-state Potts or Ising universality class.

\bibliography{refs.bib}

\end{document}